\documentclass[trackchanges]{aastex701}

\usepackage{subfigure}
\usepackage{multirow}
\usepackage{longtable}
\usepackage{amsthm,amsmath,amssymb}
\usepackage{mathrsfs}
\usepackage{rotating}
\usepackage{threeparttable}
\usepackage{graphicx} 

\graphicspath{{./}{fig/}}

\begin{document}

\title{Fast-Cooling Synchrotron in Decaying Magnetic Fields: Implications for the GRB Spectral Distribution}

\correspondingauthor{Li Zhang}\email{lizhang@ynu.edu.cn}
\author[0000-0001-5681-6939]{Jia-Ming Chen}
\affiliation{Department of Astronomy, School of Physics and Astronomy, Key Laboratory of Astroparticle Physics of Yunnan Province, Yunnan University, Kunming 650091,People's Republic of China}
\email{chenjiaming5821@163.com}  

\author[0000-0002-3132-1507]{Ke-Rui Zhu}
\affiliation{Department of Astronomy, School of Physics and Astronomy, Key Laboratory of Astroparticle Physics of Yunnan Province, Yunnan University, Kunming 650091,People's Republic of China}
\email{zhukerui0810@163.com}

\author[0000-0003-3846-0988]{Zhao-Yang Peng}
\affiliation{College of Physics and Electronics information, Yunnan Normal University, Kunming 650500, People's Republic of China}
\email{pengzhaoyang412@163.com}

\author[0000-0003-0170-9065]{Yong-Gang Zheng}
\affiliation{Department of Physics, Yunnan Normal University, Kunming 650500, People's Republic of China}
\email{ynzyg@ynu.edu.cn}

\author{Yun-Lu Gong}
\affiliation{Department of Astronomy, School of Physics and Astronomy, Key Laboratory of Astroparticle Physics of Yunnan Province, Yunnan University, Kunming 650091,People's Republic of China}
\email{gyunlu2021@163.com}

\author{Shan Chang}
\affiliation{Department of Astronomy, School of Physics and Astronomy, Key Laboratory of Astroparticle Physics of Yunnan Province, Yunnan University, Kunming 650091,People's Republic of China}
\email{changshan@ynu.edu.cn}

\author{Shi-Ting Tian}
\affiliation{Department of Astronomy, School of Physics and Astronomy, Key Laboratory of Astroparticle Physics of Yunnan Province, Yunnan University, Kunming 650091,People's Republic of China}
\email{tianshiting2021@163.com}

\author[0000-0002-5880-8497]{Li Zhang}
\affiliation{Department of Astronomy, School of Physics and Astronomy, Key Laboratory of Astroparticle Physics of Yunnan Province, Yunnan University, Kunming 650091,People's Republic of China}
\email{lizhang@ynu.edu.cn}

\begin{abstract}

\end{abstract}

\begin{abstract}
The prompt-emission spectra of gamma-ray bursts (GRBs) are commonly described by the empirical Band function. The typical low-energy spectral index is $\sim -1$, which poses a challenge to standard synchrotron radiation models. We systematically investigate a fast-cooling synchrotron model with a decaying magnetic field and test, within an observation-consistent pipeline, whether it reproduces the Band-fit parameter distributions in the GBM catalog, in a statistical sense. We solve the electron continuity equation with synchrotron, adiabatic, and synchrotron self-Compton cooling to obtain the time-dependent electron distribution and synthetic spectra; we then forward-fold through the GBM response matrices and recover $(\alpha, \beta, E_p)$ with Band fits. We find that magnetic-field decay can harden the recovered $\alpha$ relative to the fast-cooling limit in part of parameter space, but the effect is not robust and is sensitive to the location of $E_p$ within the finite band and to spectral curvature; varying key physical scales reshapes the recovered $\alpha$ distribution, indicating that catalog $\alpha$ often represents an effective in-band slope rather than the asymptotic index. SSC cooling provides modest additional hardening and, in our setups, does not stabilize $\alpha$ near the observed peak. Using Monte Carlo samples designed to mimic the observations, the model yields $\alpha$ mostly between $-1.5$ and $-0.8$, but remains centered around $\alpha \approx -1.5$. Overall, while decaying-field fast-cooling synchrotron can partially alleviate overly soft spectra expected from standard fast-cooling synchrotron emission, it still falls short of reproducing the GBM $\alpha$ distribution at the population level, implying that additional physical processes are required.
\end{abstract}


\section{Introduction} \label{intro}
Gamma-ray bursts (GRBs) are among the most energetic transients in the Universe. Despite decades of observations and modeling, the radiation mechanism during the prompt phase remains uncertain \citep{2014IJMPD..2330002Z,2015PhR...561....1K,2018pgrb.book.....Z}. Accurate characterization of the observed spectra is a key route to constraining the underlying physics. Empirically, GRB spectra are well described by the Band function, a smoothly joined broken power law with low- and high-energy spectral indices $(\alpha,\beta)$ and a peak energy $E_{\rm p}$ \citep{1993ApJ...413..281B}. While the Band function captures the principal features, its physical origin is debated.

Two broad scenarios are commonly discussed for the Band component: (i) synchrotron radiation from relativistic electrons accelerated by shocks or magnetic reconnection in an optically thin region  \citep{1994ApJ...432..181M,1998MNRAS.296..275D,2000ApJ...543..722L,2011A&A...526A.110D,2011ApJ...726...90Z,2020NatAs...4..174B}, and (ii) Comptonization of quasi-thermal photospheric photons \citep{2000ApJ...530..292M,2010ApJ...725.1137L,2010MNRAS.407.1033B}. Both can reproduce the observed ranges of $E_{\rm p}$ and $\beta$, complicating discrimination. Classical fast-cooling synchrotron (FCS) predicts $\alpha=-3/2$, whereas slow-cooling synchrotron (SCS) yields $\alpha=-2/3$ 
\citep{1998ApJ...497L..17S}. Catalog studies show the observed $\alpha$ distribution peaks near $-1$ with substantial dispersion \citep{2000ApJS..126...19P,2006ApJS..166..298K,2012ApJS..199...19G}; about one third of Band-fit $\alpha$ values are inconsistent with SCS, and nearly all with FCS, leading to the ``death-line" problem \citep{1998ApJ...506L..23P,2012ApJS..199...19G}. By contrast, photospheric models tend to be too hard: in the Rayleigh--Jeans regime, a blackbody gives $\alpha\sim+1$, softening to $\alpha\sim+0.4$ with relativistic effects \citep{2010MNRAS.407.1033B,2013MNRAS.428.2430L,2014ApJ...785..112D}, still harder than typical data.

Beyond the dominant Band component, observations by Fermi and other satellites have further revealed additional spectral components, such as a quasi-thermal feature that can be identified in some bursts and an extra power-law component extending to high energies \citep{2009ApJ...706L.138A,2010ApJ...709L.172R,2013ApJ...770...32G,2021ApJ...922..255T}. \cite{2011ApJ...730..141Z} proposed that GRB spectra may be composed of the Band component, a quasi-thermal component, and a high-energy component in different combinations; the observed spectrum may arise from the superposition of any two of them, or even from the coexistence of all three. A common approach augments Band with a blackbody to reconcile synchrotron expectations with the observed $\alpha$ distribution in GBM data \citep{2013MNRAS.432.3237G,2017ApJ...846..138G,2022ApJ...932...25C}. In some cases, adding a blackbody shifts the Band-only $\alpha$ toward synchrotron values; e.g., \cite{2019ApJS..245....7L} showed that a thermal component can significantly impact parameter inference and physical interpretation. However, such conclusions depend on how well Band approximates a true synchrotron spectrum over the instrument bandpass. \cite{2015MNRAS.451.1511B} forward-folded the theoretical spectra of  synchrotron emission and ``synchrotron + blackbody" radiation through the GBM detector response, fitting them with the Band model and the ``Band + blackbody" model, respectively. The results showed that even with the inclusion of a blackbody component, the overall distribution of $\alpha$ observed could not be fully explained. This suggests that additional radiation mechanisms may need to be introduced in most GRBs.

Nevertheless, electron synchrotron radiation remains one of the important candidate radiation mechanisms. \cite{2020NatAs...4..174B} conducted a direct comparison between theoretical models and observational data, indicating that synchrotron spectra generated by electrons in the evolution from fast cooling to slow cooling can fit approximately 95\% of the time-resolved spectra from the brightest long GRB observed by Fermi-GBM. To explain the observed phenomenon of $\alpha$ exceeding the synchrotron ``line of death" , researchers have proposed various modified synchrotron models. For example, the Klein-Nishina (KN) effect on inverse Compton cooling can alter the cooling rate of fast-cooling electrons, producing a harder low-energy spectrum and thereby being consistent with a subset of observed GRB prompt spectra \citep{2009ApJ...703..675N,2009ApJ...698L..98W,2011A&A...526A.110D}. Furthermore, incorporating adiabatic losses and synchrotron losses in a decaying magnetic field can harden the fast-cooling synchrotron spectrum at low energies relative to the standard constant-field fast-cooling prediction, thereby offering a plausible explanation for the typically observed Band-like low-energy slopes \citep{2014NatPh..10..351U,2014ApJ...780...12Z,2016ApJ...816...72Z,2025A&A...693A.320D}.
 \cite{2018ApJS..234....3G}, by comprehensively considering the KN effect and large-scale decaying magnetic fields, reached similar conclusions. Some researchers have attempted to fit actual observational data using such models, but these models still require further testing with more observational data.

Here we examine Band-function fits to spectra whose intrinsic emission is synchrotron radiation in a decaying magnetic field. Following \cite{2015MNRAS.451.1511B}, we synthesize such photon spectra, forward-fold them through the Fermi-GBM response, and fit the resulting count spectra with  Band model. We assess whether the recovered parameter distributions are self-consistent with the underlying assumptions. The paper is organized as follows. Section \ref{section2} describes the physical model and numerical method. Section \ref{section3} outlines the forward-folding and Band-recovery pipeline for catalog-consistent comparisons with Fermi-GBM. Section \ref{section4} presents the main results, and Section \ref{section5} summarizes the conclusions and discusses their implications.

\section{GRB Jet Model and Numerical Method \label{section2}}

In this work, we consider a spherically symmetric thin shell that expands with a constant Lorentz factor $\Gamma = [1/(1-\beta^{2})]^{1/2}$,
to describe the dynamical evolution of the GRB emission region, where $\beta$ is the dimensionless shell velocity. The shell is assumed to remain approximately spherical during the expansion, and its comoving thickness is much smaller than the radius. In the comoving frame, the magnetic field is dominated by the toroidal component and decays monotonically with radius as \citep{2001A&A...369..694S,2014NatPh..10..351U}
\begin{equation}
B' = B_0 \left(\frac{R}{R_0}\right)^{-a},
\end{equation}
where $R = R_0 + \beta c \Gamma t'$ is the jet radius and $t'$ is the comoving time. Here $R_0$ is the characteristic radius at which the shell starts to radiate, $B_0$ is the magnetic-field strength at $R_0$, and $a$ is the power–law decay index of the magnetic field. The observer time $t_{\rm obs}$ is related to the comoving time by
\begin{equation}
{\rm d}t_{\rm obs} \simeq \frac{1+z}{2\Gamma}\,{\rm d}t',
\end{equation}
where $z$ is the redshift of the burst.

Electrons in the shell are continuously accelerated by internal shocks \citep{1994ApJ...432..181M} or magnetic reconnection \citep{2011ApJ...726...90Z} and form a nonthermal high–energy population. For simplicity, we assume that electrons are injected in the comoving frame with a power-law distribution in Lorentz factor,
\begin{equation}
Q'(\gamma'_e) =
\begin{cases}
Q_0\,\gamma_e^{\prime -p}, & \gamma'_{e,m} < \gamma'_e < \gamma'_{e,\max},\\[4pt]
0, & \text{otherwise},
\end{cases}
\end{equation}
where $Q_0$ is the injection rate, $p$ is the injection index, and $\gamma'_{e,m}$ and $\gamma'_{e,\max}$ are the minimum and maximum Lorentz factors of the injected electrons, respectively.

The injected electrons lose energy through several cooling processes, including synchrotron (SYN), adiabatic (ADI), and synchrotron self–Compton (SSC) scattering. The evolution of the electron distribution in time and energy is governed by the kinetic continuity equation in the comoving frame \citep{2011hea..book.....L,2014NatPh..10..351U,2014ApJ...780...12Z,2018ApJS..234....3G},
\begin{equation}
\frac{\partial}{\partial t'}\left(\frac{{\rm d}N'_e}{{\rm d}\gamma'_e}\right)
 + \frac{\partial}{\partial \gamma'_e}\left[\dot{\gamma}'_{e,\rm tot}
 \left(\frac{{\rm d}N'_e}{{\rm d}\gamma'_e}\right)\right]
 = Q'(\gamma'_e),
\label{eq4}
\end{equation}
where ${\rm d}N'_e/{\rm d}\gamma'_e$ is the differential number of electrons per unit Lorentz factor, and
\begin{equation}
\dot{\gamma}'_{e,\rm tot} = \dot{\gamma}'_{e,\rm syn}
 + \dot{\gamma}'_{e,\rm adi}
 + \dot{\gamma}'_{e,\rm SSC}
\end{equation}
is the total cooling rate.

For an electron with Lorentz factor $\gamma'_e$, the synchrotron cooling rate is given by \citep{1979rpa..book.....R}
\begin{equation}
\dot{\gamma}'_{e,\rm syn} =
 -\frac{\sigma_T B^{\prime 2}\gamma_e^{\prime 2}}{6\pi m_e c},
\end{equation}
where $\sigma_T$ is the Thomson cross section and $m_e$ is the electron mass. The adiabatic cooling rate, which reflects the energy loss due to the radial expansion of the shell, is given by \citep{2012ApJ...761..147U,2018ApJS..234....3G,2020MNRAS.498.3492C}
\begin{equation}
\dot{\gamma}'_{e,\rm adi}
 = -\frac{2\gamma'_e}{3R}\frac{{\rm d}R}{{\rm d}t'}
 = -\frac{2}{3}\frac{\beta c\Gamma\,\gamma'_e}{R}.
\end{equation}

The SSC cooling rate can be written as \citep{2008ApJ...686..181F,2016MNRAS.459.3175Y,2020MNRAS.498.3492C}
\begin{equation}
\dot{\gamma}'_{e,\rm SSC}
  = -\frac{3\sigma_T m_e c^{3}}{8h^{2}}
    \int_{0}^{\infty} \frac{u'(\nu')}{\nu^{\prime 2}}
    \,G(\gamma'_e,\nu')\,{\rm d}\nu',
\end{equation}
where $h$ is Planck’s constant, $u'(\nu')$ is the comoving photon energy density per unit frequency, and
$E = \gamma'_e h\nu'/{m_e c^{2}}$ is the dimensionless parameter characterizing the typical energy scale of electron–photon scattering. The function $G(\gamma'_e,\nu')$ is the scattering kernel including Klein–Nishina corrections. Its analytic form is expressed as a piecewise function:
\begin{equation}
G(\gamma'_e,\nu') =
\begin{cases}
\displaystyle
\frac{8}{3}\,\frac{E(1+5E)}{(1+4E)^2}
-\frac{4E}{1+4E}\!\left(\frac{2}{3}+\frac{1}{2E}+\frac{1}{8E^2}\right)
+\ln(1+4E)\!\left[1+\frac{3}{E}+\frac{3}{4E^2}
+\frac{\ln(1+4E)}{2E}-\frac{\ln(4E)}{E}\right] \\[6pt]
\displaystyle\quad
-\frac{5}{2E}
+\frac{1}{E}\sum_{n=1}^{\infty}\frac{(1+4E)^{-n}}{n^{2}}
-\frac{\pi^{2}}{6E}-2, & E>1/4,\\[10pt]
\displaystyle
E^{2}\!\left(\frac{32}{9}-\frac{112}{5}E+\frac{3136}{25}E^{2}\right),
& E\le 1/4,
\end{cases}
\end{equation}
which gives the exact Klein–Nishina form for $E>1/4$ and a polynomial approximation for $E\le1/4$ to facilitate numerical evaluation. With this definition, $G(\gamma'_e,\nu')$ provides a unified description of SSC cooling from the Thomson to the Klein–Nishina regime.

The photon energy density $u'(\nu')$ is built up by the synchrotron emission of all electrons in the shell and can be written as
\begin{equation}
u'(\nu') = \frac{\Delta R'}{c}
\int_{\gamma'_{e,\min}}^{\gamma'_{e,\max}} P'(\gamma'_e,\nu')
\frac{{\rm d}N'_e}{{\rm d}\gamma'_e}\frac{1}{V'}\,{\rm d}\gamma'_e,
\end{equation}
where $\Delta R' \approx R/\Gamma$ is the shell thickness in the comoving frame and $V' = 4\pi R^{2}\Delta R'$ is the comoving shell volume. The single–electron synchrotron power at frequency $\nu'$ is \citep{1979rpa..book.....R}
\begin{equation}
P'(\gamma'_e,\nu') =
 \frac{\sqrt{3}\,q_e^{3} B' \sin\alpha'}{m_e c^{2}}
 \left[x\int_{x}^{\infty} K_{5/3}(\xi)\,{\rm d}\xi\right],
\end{equation}
where $q_e$ is the elementary charge, $\alpha'$ is the pitch angle between the electron velocity and the local magnetic field, and $K_{5/3}$ is the modified Bessel function of the second kind. Here $x = \nu'/\nu'_c$, $
\nu'_c = 3q_e B'\sin\alpha'\,\gamma_e^{\prime 2}/{4\pi m_e c}$
is the characteristic synchrotron frequency.

Neglecting equal–arrival–time–surface effects and any angular structure inside the shell, the observed specific flux can be approximated as
\begin{equation}
F_\nu \simeq \frac{\Gamma(1+z)}{2\pi D_L^{2}}
\int_{\gamma'_{e,m}}^{\gamma'_{e,\max}} P'(\gamma'_e,\nu')
\frac{{\rm d}N'_e}{{\rm d}\gamma'_e}\,{\rm d}\gamma'_e,
\end{equation}
where $D_L$ is the luminosity distance to the source and $\nu' = (1+z)\nu_{\rm obs}/\delta$ is the comoving frequency. We adopt a standard $\Lambda$CDM cosmology with
$H_{0}=67.4~\mathrm{km~s^{-1}~Mpc^{-1}}$,
$\Omega_{\rm M}=0.315$, and $\Omega_{\Lambda}=0.685$ \citep{2020A&A...641A...6P}.
Throughout this work, all simulations are performed at a fixed redshift of $z=1$. Under this setup, the cosmological choice mainly determines $D_L$ and hence the overall flux normalization, but is not expected to introduce significant relative changes in the recovered spectral shape across the simulated sample. The Doppler factor $\delta$ is approximated as $\delta \simeq \Gamma$ for on–axis viewing within a narrow jet opening angle. This expression gives the instantaneous spectrum at any observer time $t_{\rm obs}$ for a given electron distribution.

In the numerical calculations, we fix the minimum and maximum electron
Lorentz factors at $\gamma_{\min}=1$ and $\gamma_{\max}=10^{8}$,
respectively. We solve the above continuity equation numerically to obtain the time–dependent electron spectra, and then compute the corresponding radiation spectra by inserting the solutions into the radiative transfer formula. The resulting observed flux is written as
\begin{equation}
F_{\nu,{\rm obs}} = F_{\nu,{\rm obs}}
\bigl(t_{\rm obs}, E_{obs}, B_0, a, \Gamma, \gamma_m, p, R_0, Q_0, z\bigr).
\end{equation}

\begin{figure}[htbp]
\centering
\subfigure[]
{
\begin{minipage}[t]{0.47\textwidth}
\centering
\includegraphics[width=\textwidth]{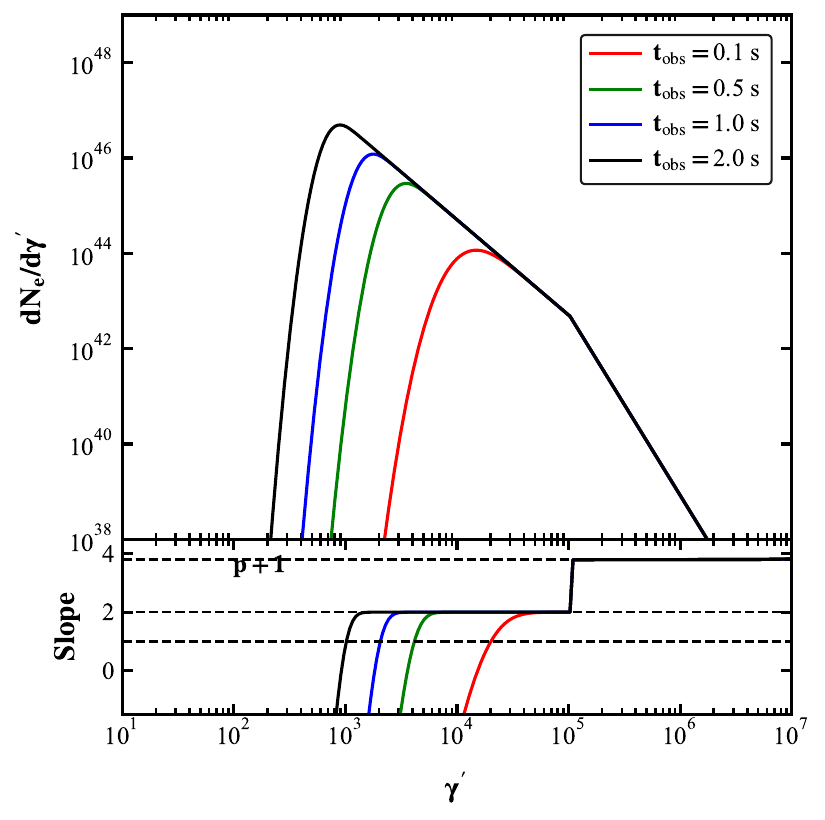}
\end{minipage}
}
\subfigure[]
{
\begin{minipage}[t]{0.47\textwidth}
\centering
\includegraphics[width=\textwidth]{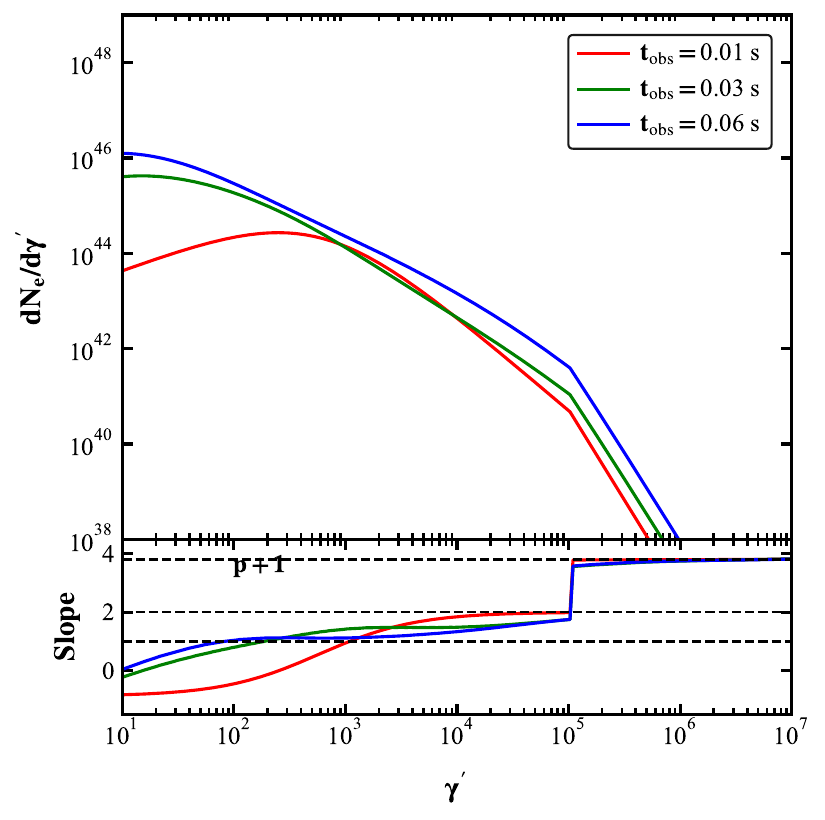}
\end{minipage}
}

\caption{Time evolution of the electron energy distribution and the corresponding local spectral slope.
The upper panels show the comoving electron distribution $dN_e/d\gamma'$ as a function of the Lorentz factor $\gamma$,
and the lower panels show the corresponding power-law index (Slope).
Panel (a) corresponds to the M1 model in Geng18, and panel (b) corresponds to the M3 model.
Curves of different colors represent different observer times $t_{\rm obs}$,
and the dashed lines indicate the theoretical asymptotic slopes.\label{fig1}}
\end{figure}

We solved Equation \ref{eq4} using a fully implicit finite-difference scheme \citep{CHANG19701} and implemented the numerical code in Python to compute the electron spectrum and the corresponding radiation spectrum. To validate the code, we selected the M1 and M3 models in \cite{2018ApJS..234....3G}, adopted the same parameter settings, and compared our results with the electron and radiation spectra reported there. As shown in Figure \ref{fig1}, M1 corresponds to the standard fast-cooling synchrotron case, yielding the expected electron distribution and spectral shape, whereas M3, which includes magnetic field decay and SSC cooling, can produce a harder low-energy spectrum. Based on this code, we can further investigate how the electron and radiation spectra evolve under different parameter choices.

\section{Simulations}
\label{section3}

As described in Section~\ref{section2}, we have specified the jet dynamics and the treatment of magnetic-field decay, electron injection, and the three cooling channels. Building on that framework, we now construct a concrete simulation pipeline to generate synthetic photon spectra and fit them with the empirical Band function.

Our overall strategy follows that of \cite{2015MNRAS.451.1511B}. We first use the synchrotron model introduced in Section~\ref{section2} to compute a set of intrinsic photon spectra. These spectra are then folded through the Fermi-GBM detector response to obtain simulated count spectra. Finally, we fit the synthetic ``observations'' with the Band function to recover the spectral parameters $\alpha$, $\beta$, and $E_{\rm p}$ in exactly the same way as for real data. In contrast to \cite{2015MNRAS.451.1511B}, who focused on fast- and slow-cooling synchrotron emission in a constant magnetic field, we place particular emphasis on the fast-cooling synchrotron scenario in a magnetic field that decays as a power law with radius. This allows us to examine whether the Band parameters inferred after including magnetic-field evolution remain self-consistent with the physical picture and with the parameter distributions observed in GBM GRBs. The generation of photon spectra, forward-folding through the detector response, and spectral fitting are all performed with the \texttt{threeML} software package \citep{2015arXiv150708343V}.

In our implementation, we first calibrate the $\nu F_\nu$ peak of the synchrotron spectrum using the scaling $E_{\rm p}\propto \Gamma B\gamma_m^{2}$, where $\gamma_m$ is the minimum injection Lorentz factor and $B$ is the magnetic-field strength. Here we fix $\Gamma$ and $\gamma_m$, and choose an appropriate $B$ so that $E_{\rm p}$ reaches the desired value. Following \cite{2014NatPh..10..351U}, we constrain $B_0$ to the range $10$--$100~{\rm G}$, representative of typical magnetic-field strengths in the prompt emission region. In all simulations, the electron injection index is fixed at $p=2.8$ to reproduce the commonly observed high-energy index $\beta \approx -2.4$ of the Band function.

\begin{figure}[htbp]
\centering
\begin{minipage}[t]{0.6\textwidth}
\centering
\includegraphics[width=\textwidth]{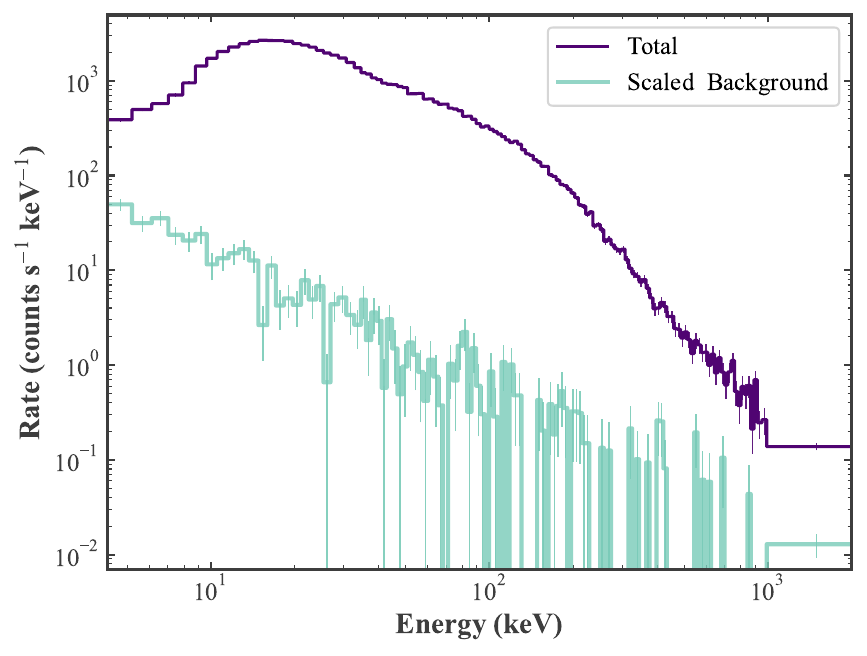}
\end{minipage}
\caption{Simulated photon counting spectrum.  \label{fig2}}
\end{figure}

Once the intrinsic photon spectra are obtained, we use the GBM detector response to generate simulated count spectra for two NaI detectors (8--1000~keV) and one BGO detector (200--40000~keV). For each simulated source spectrum, we add a synthetic background spectrum whose photon flux follows a declining power law in energy with an index of 1.5. The procedure for constructing such spectra is described in detail by \cite{2014MNRAS.445.2589B}. Figure \ref{fig2} shows the simulated photon counting spectrum. For each set of physical parameters, we generate 1000 independent realizations of the spectrum, fit each of them with the Band function, and record the resulting $\alpha$, $\beta$, and $E_{\rm p}$ values.

To investigate how the recovered Band parameters depend on the dynamical and microphysical inputs, we consider a set of representative models that explore variations in the observer time $t_{\rm obs}$, the bulk Lorentz factor $\Gamma$, the minimum injection Lorentz factor $\gamma_m$, the magnetic-field decay index $a$, and the emission onset radius $R_0$. We also control whether adiabatic cooling and SSC cooling are included in the electron energy-loss term. The parameter combinations used in this work are summarized in Table~\ref{tab:models}. 

\begin{table}
\centering
\caption{Representative simulation models used in this work.}
\label{tab:models}
\begin{tabular}{ccccccc}
\hline\hline
$t_{\rm obs}$ (s) & $\Gamma$ & $\gamma_m$ & $a$ & $R_0$ (cm) & Adiabatic & SSC \\
\hline
1.0 & 300 & $10^{5}$ & 1.0 & $10^{15}$ & Yes & No \\
1.0 & 300 & $10^{5}$ & 1.5 & $10^{15}$ & Yes & No \\
1.0 & 300 & $10^{5}$ & 1.0 & $10^{14}$ & Yes & No \\
1.0 & 300 & $10^{5}$ & 1.5 & $10^{14}$ & Yes & No \\
1.0 & 300 & $10^{4}$ & 1.0 & $10^{15}$ & Yes & No \\
1.0 & 300 & $10^{4}$ & 1.5 & $10^{15}$ & Yes & No \\
3.0 & 300 & $10^{5}$ & 1.0 & $10^{15}$ & Yes & No \\
3.0 & 300 & $10^{5}$ & 1.5 & $10^{15}$ & Yes & No \\
0.1 & 300 & $10^{5}$ & 1.0 & $10^{15}$ & Yes & Yes \\
0.5 & 300 & $10^{5}$ & 1.5 & $10^{15}$ & Yes & Yes \\
0.1 & 300 & $10^{5}$ & 0.0 & $10^{15}$ & Yes & Yes \\
\hline
\end{tabular}
\end{table}


\section{Simulation results}
\label{section4}
In this section, we systematically present the synthetic photon spectra obtained under different parameter configurations and the distributions of their Band fitting parameters, comparing them with the Fermi-GBM observational sample. We specifically investigate the influence of the emission start radius $R_0$, the minimum injection Lorentz factor $\gamma_m$, the magnetic field power-law decay index $a$, and the inclusion of SSC cooling on the statistical distribution of the low-energy spectral index $\alpha$ and the $\alpha$-$E_p$ correlation. To ensure consistency with the data processing procedure of observations, all synthetic spectra are transformed into the observational space via the forward-folding method and fitted with the Band function.

\subsection{$R_0 = 10^{15}$ cm, $\gamma_m = 10^5$: Synchrotron and Adiabatic Cooling Only}
\label{section4.1}

We fix $R_0 = 10^{15}$ cm, $\gamma_m = 10^5$, $\Gamma = 300$, and the injection normalization $Q_0 = 10^{58}$, considering only synchrotron cooling and adiabatic cooling, and scan the initial magnetic field strength $B_0$. We compare two magnetic field power-law decay indices, $a = 1.0$ and $1.5$, and perform Band fits on the synthetic spectra at two observation times, $t_{\text{obs}} = 1$ s and $3$ s, to obtain the statistical distributions of $\alpha$ and $E_p$.

\begin{figure}[htbp]
\centering
\subfigure[]
{
\begin{minipage}[t]{0.47\textwidth}
\centering
\includegraphics[width=\textwidth]{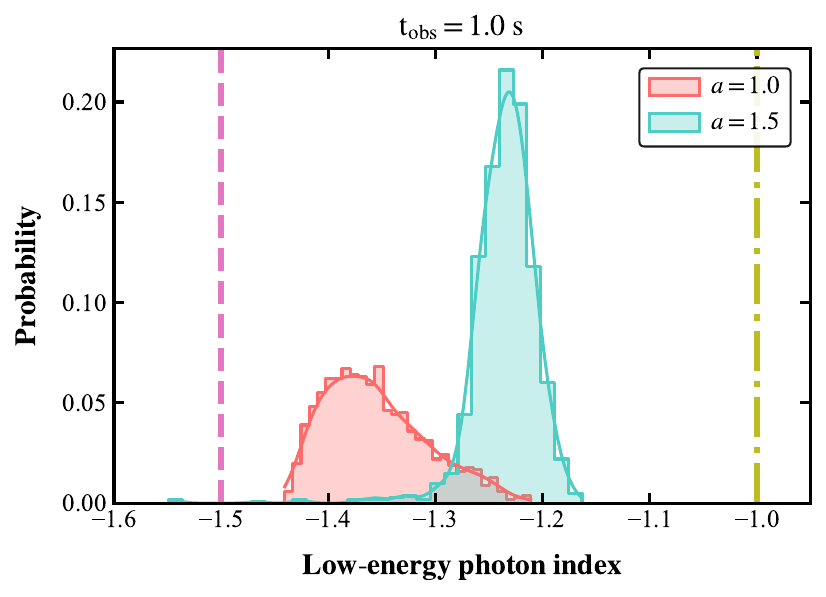}
\end{minipage}
}
\subfigure[]
{
\begin{minipage}[t]{0.47\textwidth}
\centering
\includegraphics[width=\textwidth]{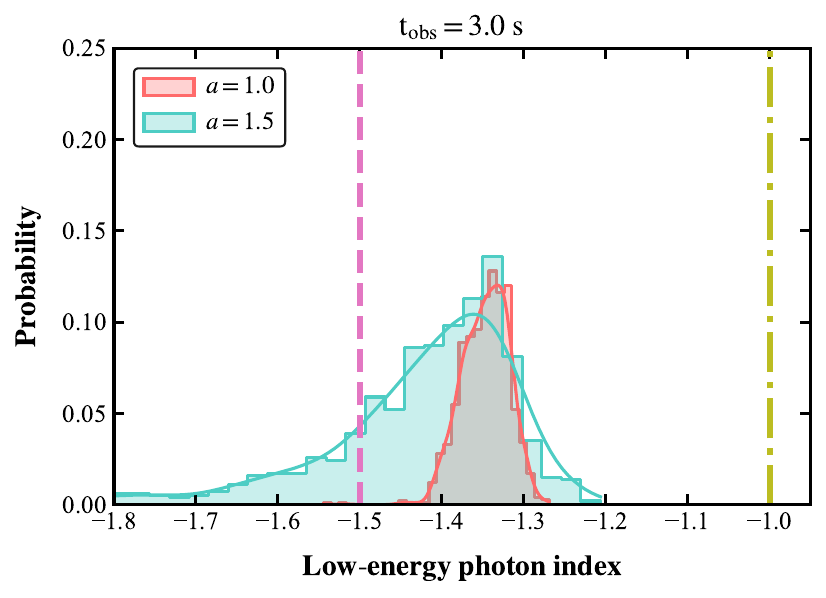}
\end{minipage}
}

\caption{Distributions of the Band low-energy spectral index $\alpha$ obtained from forward-folding and Band fitting of the synthetic spectra at (a) $t_{\rm obs}=1\,{\rm s}$ and (b) $t_{\rm obs}=3\,{\rm s}$. Results are shown for two magnetic-field decay indices, $a=1.0$ (red) and $a=1.5$ (cyan), with $R_0=10^{15}\,{\rm cm}$ and $\gamma_m=10^{5}$ (synchrotron + adiabatic cooling only). The magenta dashed line marks the synchrotron fast-cooling limit $\alpha=-3/2$, and the yellow dash--dotted line indicates $\alpha=-1$. \label{fig3}}
\end{figure}

Figure \ref{fig3} presents the distribution of $\alpha$ under different $a$ and observation times. At $t_{\text{obs}} = 1$ s, the $\alpha$ distribution for $a = 1.0$ is mainly located between $\alpha \approx -1.43$ and $-1.25$ (peaking at $\alpha \approx -1.37$), which is significantly harder than the fast-cooling limit of $-3/2$ but remains generally soft. When the magnetic field decay accelerates to $a = 1.5$, the $\alpha$ distribution shifts overall toward the hard end and narrows, concentrating mainly between $\alpha \approx -1.30$ and $-1.17$ (peaking at $\alpha \approx -1.23$). This indicates that faster magnetic field decay can effectively alleviate the too soft low-energy spectrum problem under standard fast cooling at early times. As the evolution proceeds to $t_{\text{obs}} = 3$ s, the distribution for $a = 1.0$ remains relatively concentrated and stays around $\alpha \approx -1.35$. In contrast, the distribution for $a = 1.5$ broadens significantly and exhibits a more pronounced soft tail, with some samples even yielding fitting results of $\alpha < -1.5$. This suggests that with faster magnetic field decay and at later times, spectral diversity and empirical fitting effects have a significant impact on the statistics of $\alpha$.

\begin{figure}[htbp]
\centering
\subfigure[]
{
\begin{minipage}[t]{0.50\textwidth}
\centering
\includegraphics[width=\textwidth]{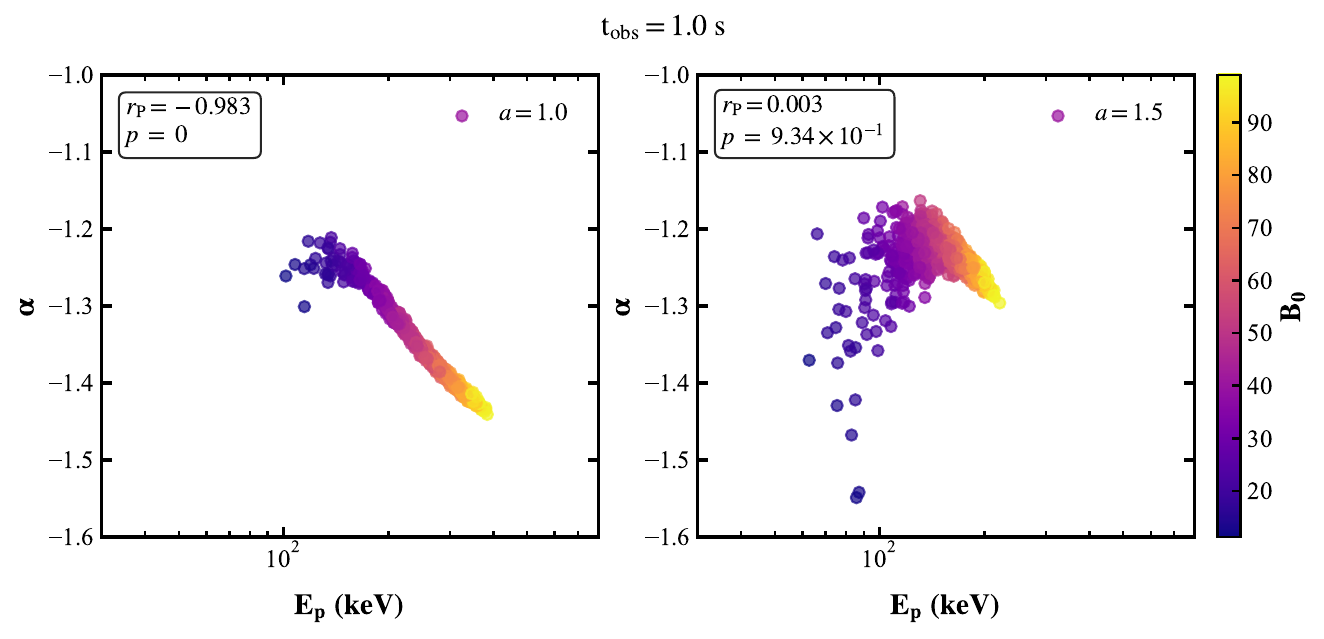}
\end{minipage}
}
\subfigure[]
{
\begin{minipage}[t]{0.50\textwidth}
\centering
\includegraphics[width=\textwidth]{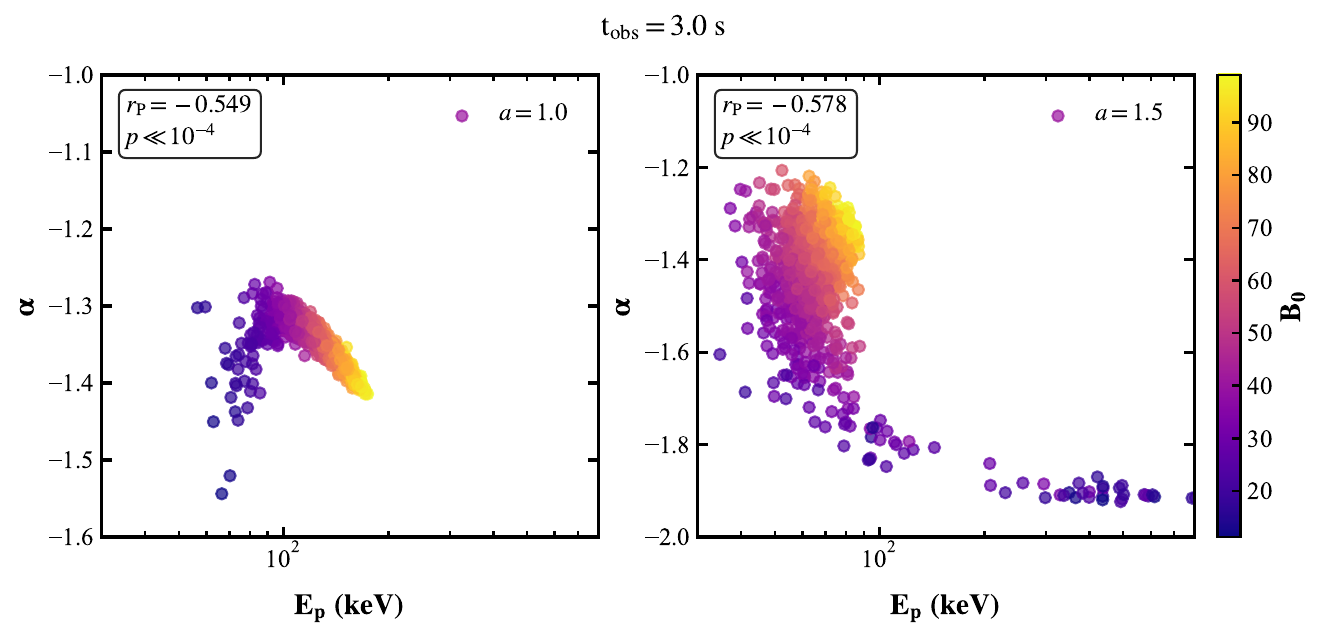}
\end{minipage}
}

\caption{Band low-energy spectral index $\alpha$ versus the fitted peak energy $E_p$ at (a) $t_{\rm obs}=1\,{\rm s}$ and (b) $t_{\rm obs}=3\,{\rm s}$. The left and right panels correspond to magnetic-field decay indices $a=1.0$ and $a=1.5$, respectively. The point color indicates the initial magnetic-field strength $B_0$. All points are obtained from forward-folding and Band fitting of the synthetic spectra for the synchrotron + adiabatic-cooling model, with $R_0=10^{15}\,{\rm cm}$ and $\gamma_m=10^{5}$. The Pearson correlation coefficient $r_{\rm P}$ between $\alpha$ and $\log_{10}(E_p/{\rm keV})$, together with the corresponding $p$-value, is shown in each panel.
 \label{fig4}}
\end{figure}

Figure \ref{fig4} shows the $\alpha$--$E_p$ relation, with the Pearson correlation coefficient given in each panel. For $a=1.0$, $\alpha$ and $E_p$ are anti-correlated at both epochs. The anti-correlation is extremely strong at $t_{\rm obs}=1.0$ s ($r_{\rm P}=-0.983$), such that larger $E_p$ values, generally associated with larger $B_0$, correspond to softer $\alpha$ values approaching $-3/2$, while smaller $E_p$ corresponds to relatively harder spectra. At $t_{\rm obs}=3.0$ s, the anti-correlation persists but is weaker ($r_{\rm P}=-0.549$) and shows larger scatter. For $a=1.5$, the correlation is much less regular. At $t_{\rm obs}=1.0$ s, the points cluster around $\alpha\approx -1.2$ to $-1.3$ with no significant monotonic trend ($r_{\rm P}=0.003$). At $t_{\rm obs}=3.0$~s, the dispersion becomes substantially larger and a soft tail extending to $\alpha\lesssim -1.8$ emerges, mainly at low $E_p$. Although a moderate anti-correlation is recovered statistically ($r_{\rm P}=-0.578$), the relation is much broader than in the $a=1.0$ cases. This indicates that, for late-time spectra with $a=1.5$, the recovered $\alpha$ is shaped not only by the peak energy itself, but also by spectral curvature and finite band fitting effects.

\begin{figure}[htbp]
\centering
\subfigure[]
{
\begin{minipage}[t]{0.47\textwidth}
\centering
\includegraphics[width=\textwidth]{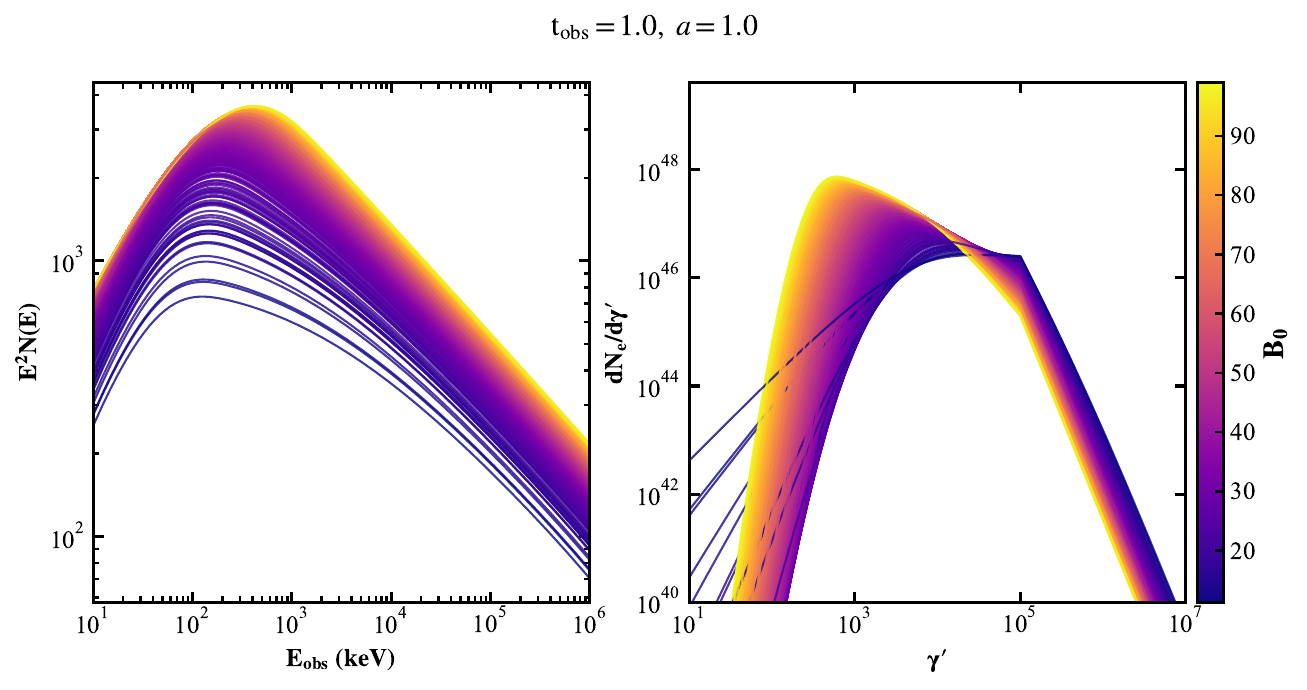}
\end{minipage}
}
\subfigure[]
{
\begin{minipage}[t]{0.47\textwidth}
\centering
\includegraphics[width=\textwidth]{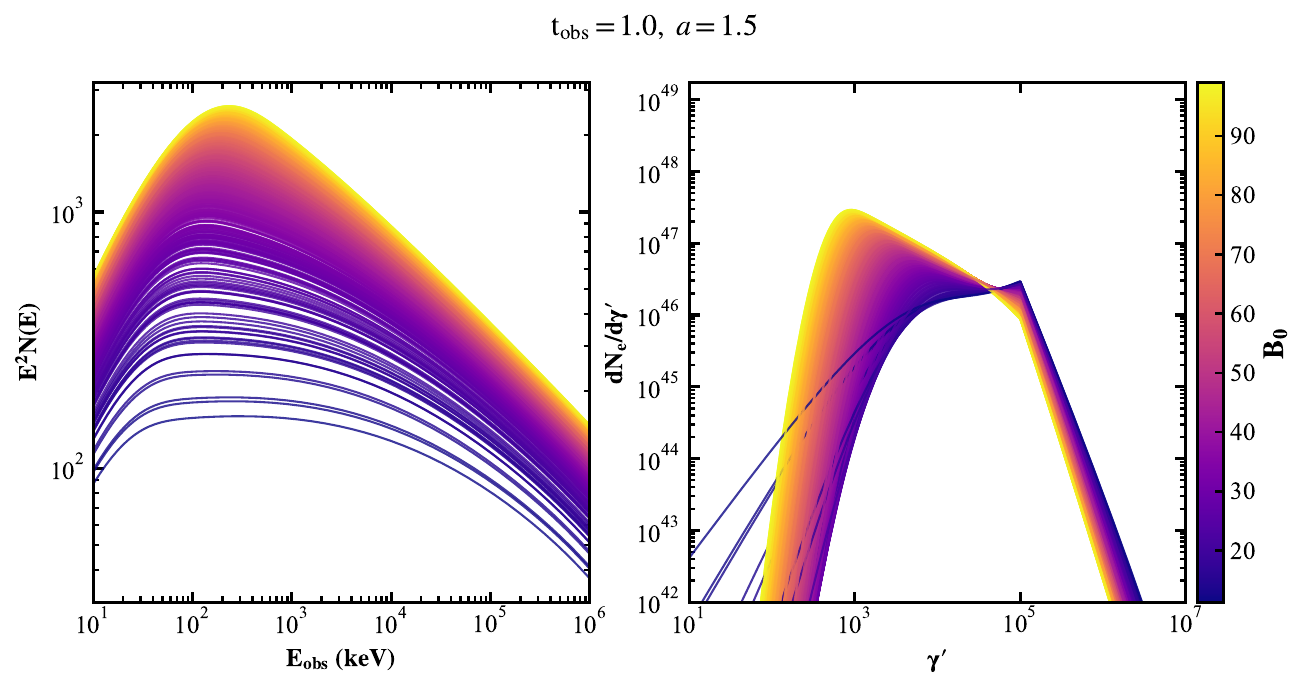}
\end{minipage}
}
\subfigure[]
{
\begin{minipage}[t]{0.47\textwidth}
\centering
\includegraphics[width=\textwidth]{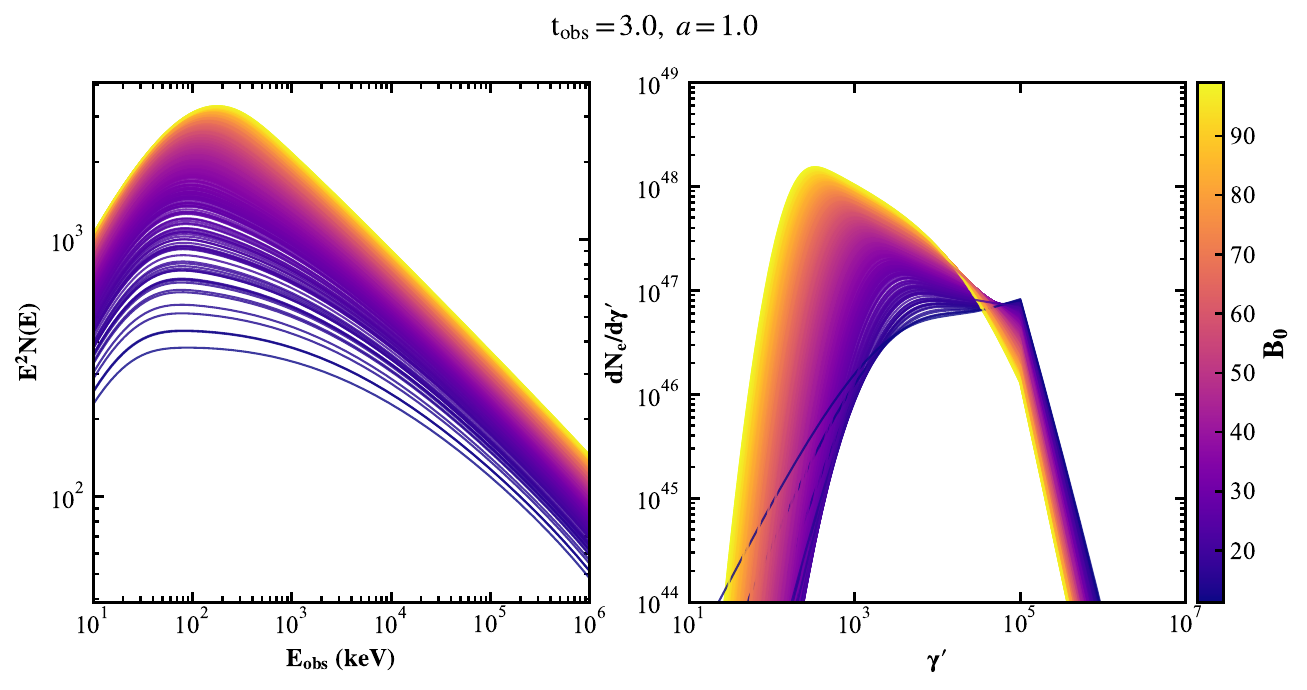}
\end{minipage}
}
\subfigure[]
{
\begin{minipage}[t]{0.47\textwidth}
\centering
\includegraphics[width=\textwidth]{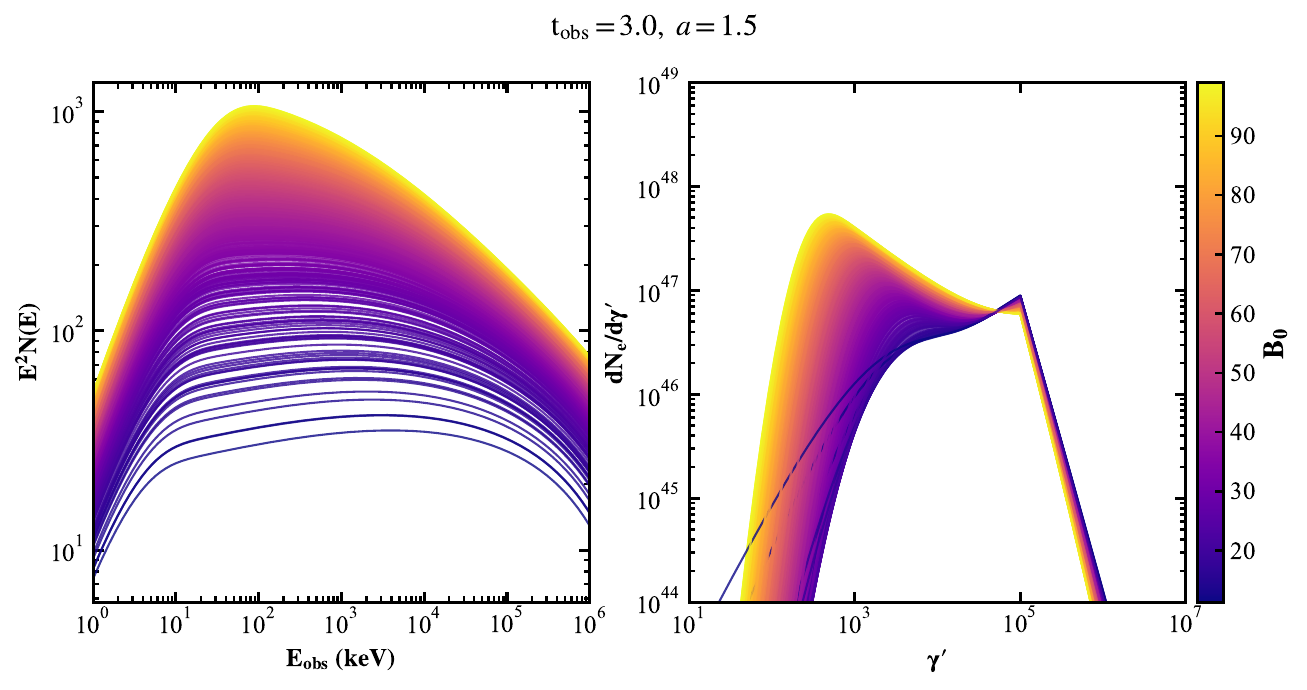}
\end{minipage}
}

\caption{Photon spectra and electron distributions for the synchrotron+adiabatic cooling  models in Section \ref{section4.1}. For each case, the left panel shows the synthetic $\nu F_\nu$ spectrum ($E^2N(E)$) in the observer frame, and the right panel shows the corresponding comoving electron distribution $dN_e/d\gamma'$. Panels (a)--(d) correspond to $(t_{\rm obs},a)=(1\,{\rm s},1.0)$, $(1\,{\rm s},1.5)$, $(3\,{\rm s},1.0)$,and $(3\,{\rm s},1.5)$, respectively. Curves are color-coded by the initial magnetic-field strength $B_0$ (see color bars). All spectra are computed with $R_0=10^{15}\,{\rm cm}$, $\gamma_m=10^{5}$, and $\Gamma=300$.
 \label{fig5}}
\end{figure}

Figure \ref{fig5} presents the instantaneous photon spectra $E^2 N(E)$ and the corresponding electron distributions $dN_e/d\gamma'$ under different $B_0$. As $B_0$ increases, $E_p$ shifts systematically upward and the spectral normalization rises, consistent with the synchrotron radiation scaling relation $E_p \propto B \gamma_m^2$ (when $\gamma_m$ is fixed). In the case of $a = 1.0$, the electron distribution forms a low-energy branch at $\gamma' \lesssim \gamma_m$ modulated jointly by synchrotron radiation and adiabatic expansion, and exhibits a clear spectral break near $\gamma' \approx \gamma_m$. As time evolves to $t_{\text{obs}} = 3$ s, the cumulative effect of cooling modulation strengthens, causing the relevant features to migrate overall toward lower $\gamma'$. In contrast, the faster magnetic field decay for $a = 1.5$ makes the system more likely to enter a weak magnetic field region at larger radii, significantly reducing the synchrotron cooling efficiency of electrons injected at late times, which gradually becomes dominated by adiabatic cooling. In the weak magnetic field phase, the characteristic spectral power of single-electron synchrotron radiation near the peak satisfies $P_{\nu, \max} \propto B$, while $\nu_m \propto B \gamma_m^2$ shifts downward. This allows the fitting energy band to more easily sample the strong curvature region near the spectral peak, resulting in the atypical spectral shapes shown by the dark curves in Figure 6. In such cases of strong curvature or downward-shifted spectral peaks, the $\alpha$ obtained from Band fitting is often merely an ``effective slope" within the limited energy band. The fitting tends to compensate for the curvature and peak offset with a softer $\alpha$, thus naturally producing results where $\alpha < -1.5$. In other words, these outliers mainly reflect the enhancement of spectral curvature and empirical fitting effects under weak magnetic field and adiabatic cooling dominated conditions, rather than the true low-energy asymptotic slope of synchrotron radiation breaking through the fast-cooling limit.

\subsection{$R_0 = 10^{14}$ cm, $\gamma_m = 10^5$: Extreme Scenario of Small Emission Radius}
\label{section4.2}

To investigate the influence of the emission radius on the results, we reduce the emission start radius to $R_0 = 10^{14}$ cm and compare the statistical properties of $a = 1.0$ and $1.5$ at $t_{\text{obs}} = 1$ s. Since the corresponding evolutionary scale ratio $R/R_0$ at the same observation time is significantly larger, the impact of the magnetic field decay $B \propto R^{-a}$ is markedly amplified, causing the model to more readily enter the phase of rapid evolution and weak magnetic fields.

\begin{figure}[htbp]
\centering
\begin{minipage}[t]{0.5\textwidth}
\centering
\includegraphics[width=\textwidth]{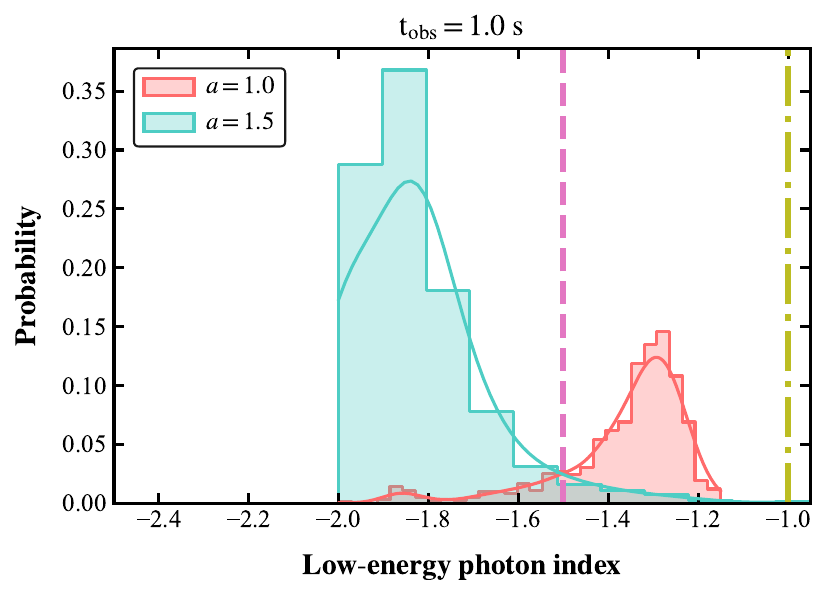}
\end{minipage}
\caption{Distribution of the Band low-energy spectral index $\alpha$ at $t_{\rm obs}=1\,\mathrm{s}$ for the $R_0=10^{14}\,\mathrm{cm}$, $\gamma_m=10^{5}$ model, obtained from forward-folding and Band fitting. The notation and line styles are the same as in Figure \ref{fig3}  \label{fig6}}
\end{figure}

\begin{figure}[htbp]
\centering
\begin{minipage}[t]{0.5\textwidth}
\centering
\includegraphics[width=\textwidth]{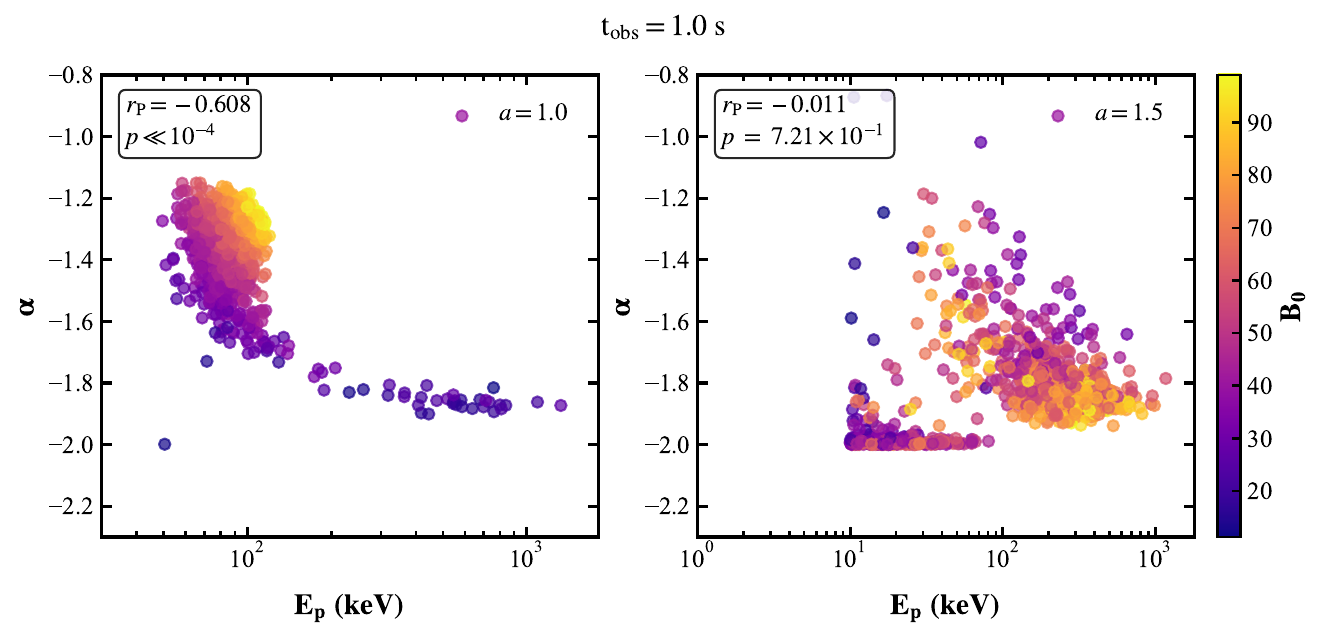}
\end{minipage}
\caption{Relation between the low-energy spectral index $\alpha$ and the peak energy $E_p$ for the $R_0=10^{14}\,\mathrm{cm}$, $\gamma_m=10^{5}$ model  at $t_{\rm obs}=1\,\mathrm{s}$. Similar to Figure \ref{fig4}.}
\label{fig7}
\end{figure}

Figure \ref{fig6} shows that for $a = 1.0$, $\alpha$ remains mainly concentrated between $-1.4$ and $-1.2$; in contrast, the results for $a = 1.5$ shift significantly toward the soft end overall and broaden considerably, with the bulk of the distribution falling between $-2.0$ and $-1.6$, yielding a substantial number of fitting values where $\alpha < -1.5$. Correspondingly, the $\alpha$--$E_p$ relation in Figure \ref{fig7} shows substantially enhanced dispersion compared with Figure~\ref{fig4}. For $a=1.0$, a moderate anti-correlation is still present ($r_{\rm P}=-0.608$), although the relation is already noticeably broader than in the previous case. For $a=1.5$, the dispersion becomes much stronger and a pronounced pile-up emerges near $\alpha\approx -2$, while the Pearson coefficient drops to $r_{\rm P}=-0.011$, indicating that the monotonic $\alpha$--$E_p$ relation has essentially disappeared. This implies that, in this parameter regime, the Band-recovered $\alpha$ is influenced more strongly by spectral peak migration, spectral curvature, and finite-band fitting effects, rather than by a stable low-energy asymptotic power-law slope.

\begin{figure}[htbp]
\centering
\subfigure[]
{
\begin{minipage}[t]{0.47\textwidth}
\centering
\includegraphics[width=\textwidth]{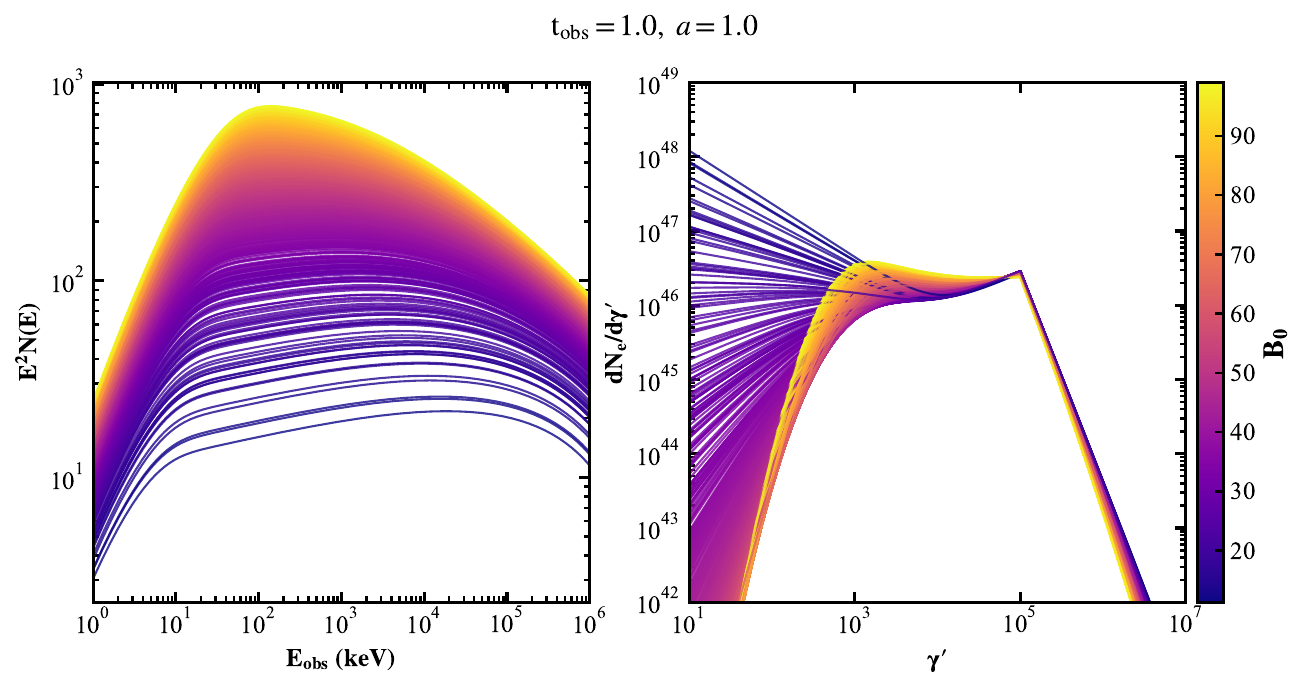}
\end{minipage}
}

\subfigure[]
{
\begin{minipage}[t]{0.47\textwidth}
\centering
\includegraphics[width=\textwidth]{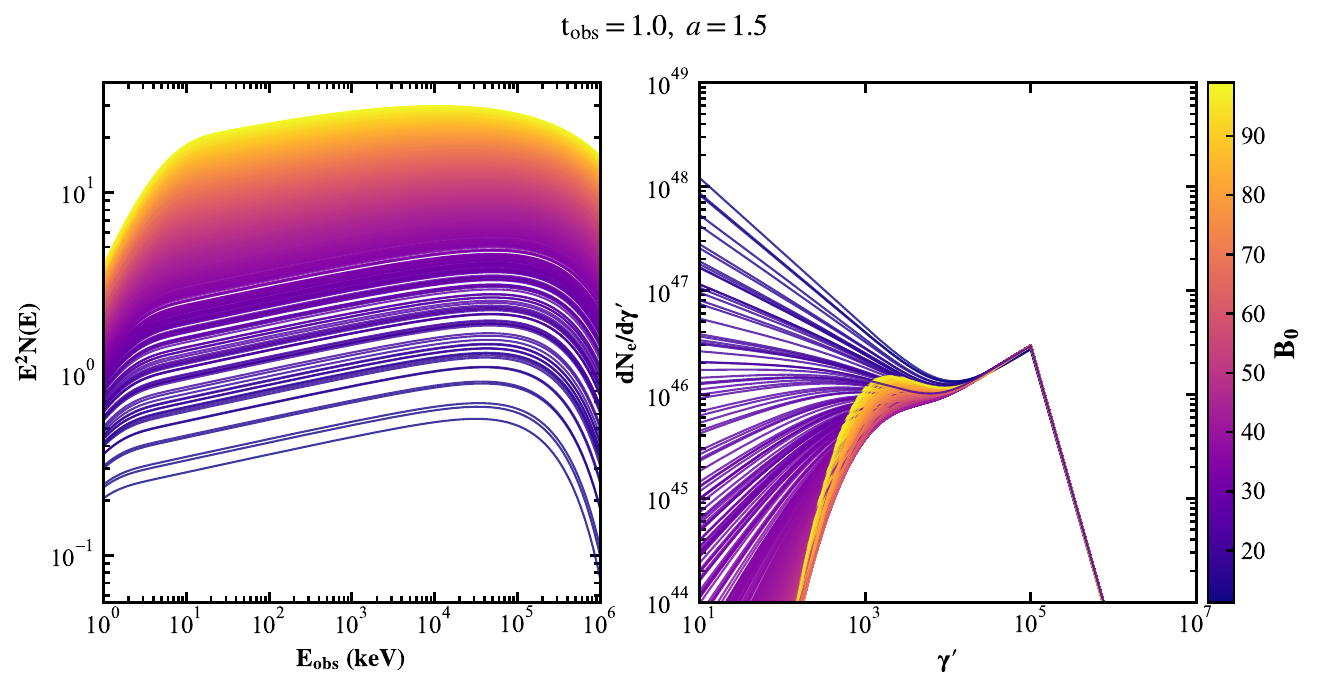}
\end{minipage}
}
\caption{Photon spectra and electron distributions for the $R_0=10^{14}\,\mathrm{cm}$, $\gamma_m=10^{5}$ models (synchrotron + adiabatic cooling only) at $t_{\rm obs}=1\,\mathrm{s}$. Similar to Figure \ref{fig5}.
 \label{fig8}}
\end{figure}

Figure \ref{fig8} further demonstrates that these ``anomalously soft" statistics are primarily driven by samples with $a = 1.5$ and smaller $B_0$. At $t_{\text{obs}} = 1$ s, the magnetic field has decayed to a lower level; the downward shift of the spectral peak is accompanied by stronger curvature, making it easier to yield softer Band effective slopes, or even $\alpha < -1.5$. The underlying physical mechanism is consistent with the ``atypical spectral shapes" observed in the weak magnetic field samples discussed in Section \ref{section4.1}.

\subsection{$R_0 = 10^{15}$ cm, $\gamma_m = 10^4$: Scenario with Lower Injection Energy}

In this subsection, we fix $R_0 = 10^{15}\,\mathrm{cm}$ and reduce the minimum injection Lorentz factor to $\gamma_m = 10^4$, while keeping the other parameters the same as in Section \ref{section4.1}. We continue to consider only synchrotron cooling and adiabatic expansion cooling. Since the synchrotron characteristic frequency satisfies $\nu_m \propto B\,\gamma_m^2$, reducing $\gamma_m$ shifts the spectral peak $E_p$ systematically toward lower energies for a given $B_0$. This changes the position of the GBM energy band relative to the spectral peak and, consequently, affects the effective low-energy spectral index $\alpha$ recovered from the Band fit.

\begin{figure}[htbp]
\centering
\begin{minipage}[t]{0.5\textwidth}
\centering
\includegraphics[width=\textwidth]{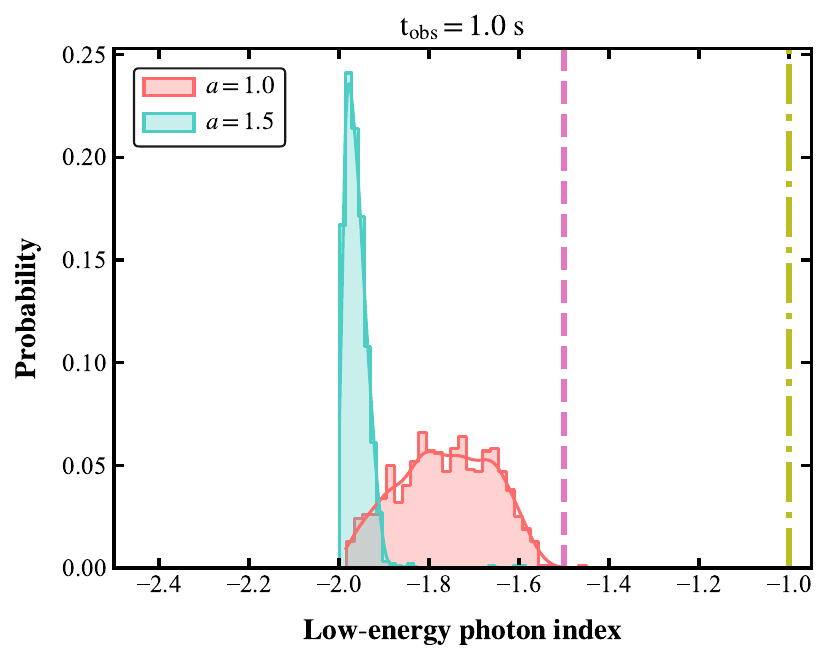}
\end{minipage}
\caption{Distribution of the Band low-energy spectral index $\alpha$ at $t_{\rm obs}=1\,\mathrm{s}$ for the $R_0=10^{15}\,\mathrm{cm}$, $\gamma_m=10^{4}$ model, obtained from forward-folding and Band fitting. \label{fig9}}
\end{figure}

\begin{figure}[htbp]
\centering
\subfigure[]
{
\begin{minipage}[t]{0.47\textwidth}
\centering
\includegraphics[width=\textwidth]{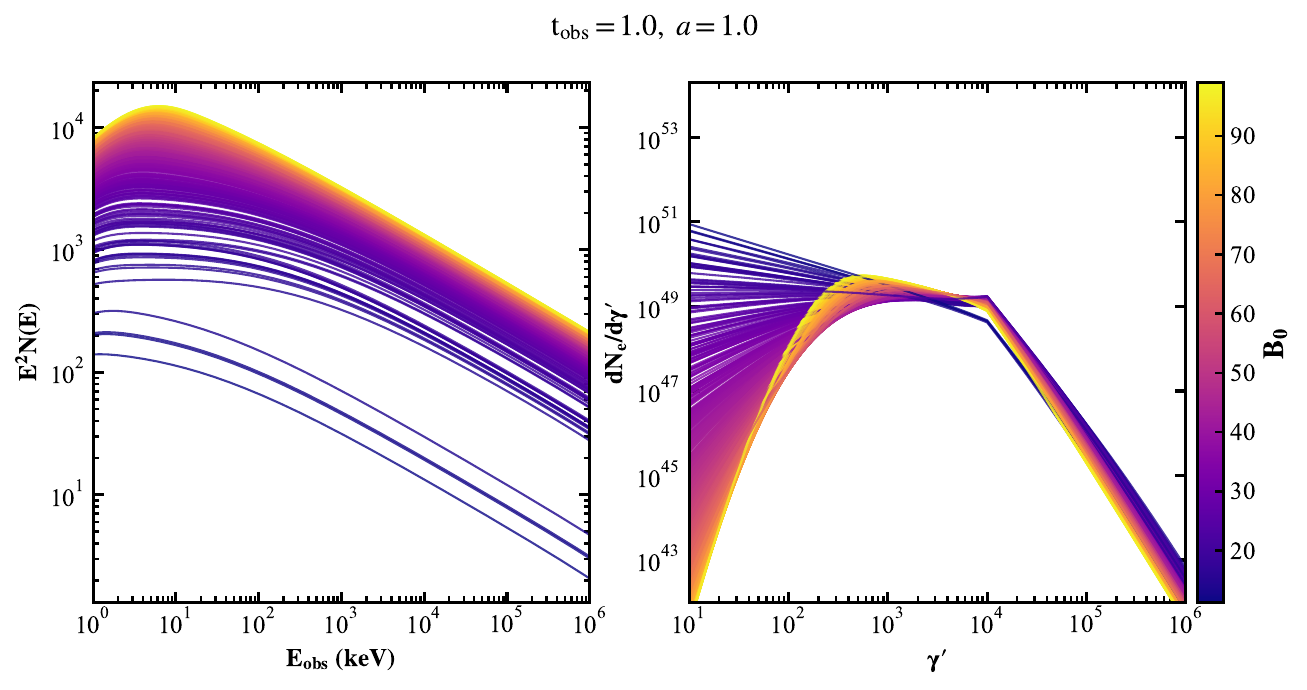}
\end{minipage}
}

\subfigure[]
{
\begin{minipage}[t]{0.47\textwidth}
\centering
\includegraphics[width=\textwidth]{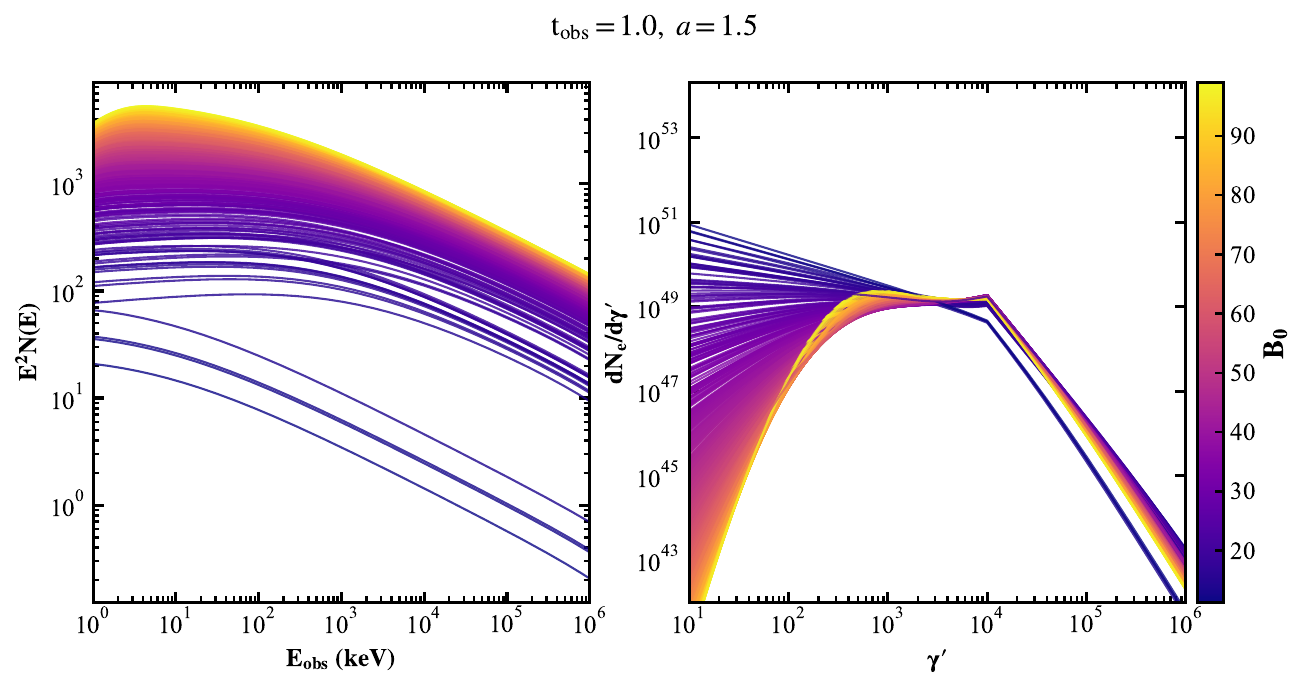}
\end{minipage}
}
\caption{Photon spectra and electron distributions for the $R_0=10^{15}\,\mathrm{cm}$, $\gamma_m=10^{4}$ models (synchrotron + adiabatic cooling only) at $t_{\rm obs}=1\,\mathrm{s}$. Similar to Figure \ref{fig5}.
 \label{fig10}}
\end{figure}

Figure \ref{fig9} shows the $\alpha$ distributions at $t_{\rm obs}=1\,\mathrm{s}$. For $a=1.0$, $\alpha$ is mainly distributed between $-1.9$ and $-1.6$, with a relatively broad spread. In contrast, for $a=1.5$, the distribution becomes much more concentrated and clusters around $\alpha \approx -2.0$, indicating a more extreme soft-spectrum behavior. Figure \ref{fig10} further presents the corresponding observed spectra and electron distributions. The reduction in $\gamma_m$ causes the spectral peak $E_p$ to shift toward lower energies. As a result, the finite bandwidth of the GBM detector no longer covers the asymptotic power-law regime of the low-energy spectrum; instead, it directly probes the spectral peak where curvature is significant, leading to a strong modulation of the fitted $\alpha$ by this spectral curvature. This effect is further enhanced by the stronger contribution from the cooling-accumulated electrons at $\gamma'_e \lesssim \gamma_m$. For $a=1.5$, the magnetic field decays more rapidly, causing the spectral peak to shift to lower energies more strongly and making the observational band more likely to probe the high-curvature region near the peak. Consequently, the fitted $\alpha$ more easily converges toward $\alpha \approx -2.0$. These results indicate that, in the low-injection-energy regime, the recovered spectral shape is controlled more strongly by the spectral-peak location and finite-band effects, and that this tendency is amplified when the magnetic field decays more rapidly.

\subsection{$R_0 = 10^{15}$ cm, $\gamma_m = 10^5$: Inclusion of SSC Cooling}
\label{section4.4}

To investigate the influence of SSC cooling on the Band fitting parameters, we incorporate the SSC cooling term and select three representative parameter combinations for comparison: $t_{\text{obs}} = 0.1$ s with $a = 0$ (constant magnetic field); $t_{\text{obs}} = 0.1$ s with $a = 1.5$ (rapidly decaying magnetic field); and $t_{\text{obs}} = 0.5$ s with $a = 1.0$ (later time, slowly decaying magnetic field), while keeping other parameters unchanged.

\begin{figure}[htbp]
\centering
\begin{minipage}[t]{0.5\textwidth}
\centering
\includegraphics[width=\textwidth]{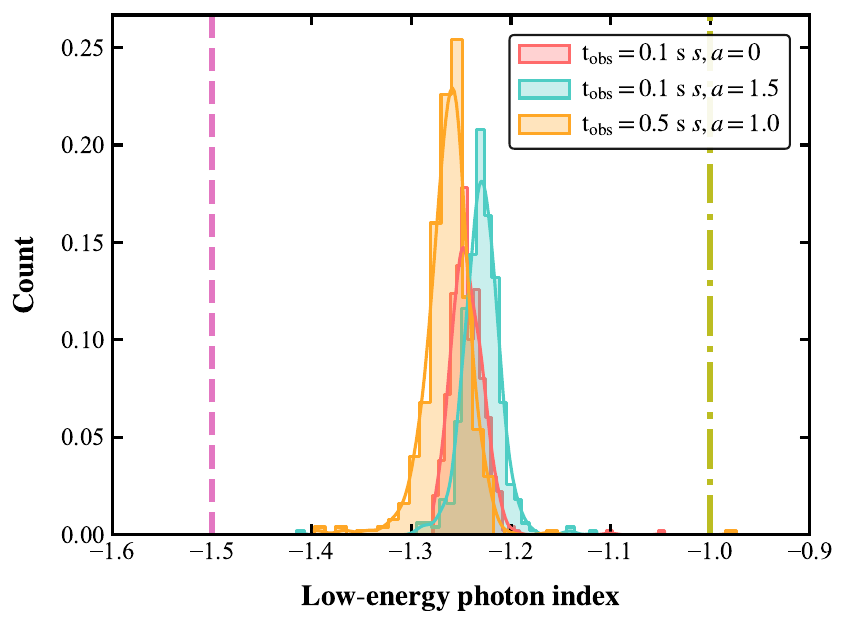}
\end{minipage}
\caption{Distributions of the Band-fitted low-energy spectral index $\alpha$ for the models incorporating SSC cooling in Section \ref{section4.4}, obtained from forward-folding and Band-model fitting. The three histograms correspond to $(t_{\rm obs}, a) = (0.1\,\mathrm{s}, 0)$ (constant magnetic field), $(0.1\,\mathrm{s}, 1.5)$ (rapidly decaying magnetic field), and $(0.5\,\mathrm{s}, 1.0)$ (later evolution with moderate decay), as indicated in the legend.
\label{fig11}}
\end{figure}

Figure \ref{fig11} shows that the $\alpha$ distributions for all three SSC model sets exhibit narrow peaks. For $a = 0$, the peak is located at $\alpha \approx -1.25$; after introducing the decaying magnetic field ($a = 1.5$), $\alpha$ shifts slightly toward the hard end (peak at $\alpha \approx -1.22$). When evolving to $t_{\text{obs}} = 0.5$ s ($a = 1.0$), the $\alpha$ distribution shifts slightly overall toward the soft end and broadens marginally.

\begin{figure}[htbp]
\centering
\begin{minipage}[t]{\textwidth}
\centering
\includegraphics[width=\textwidth]{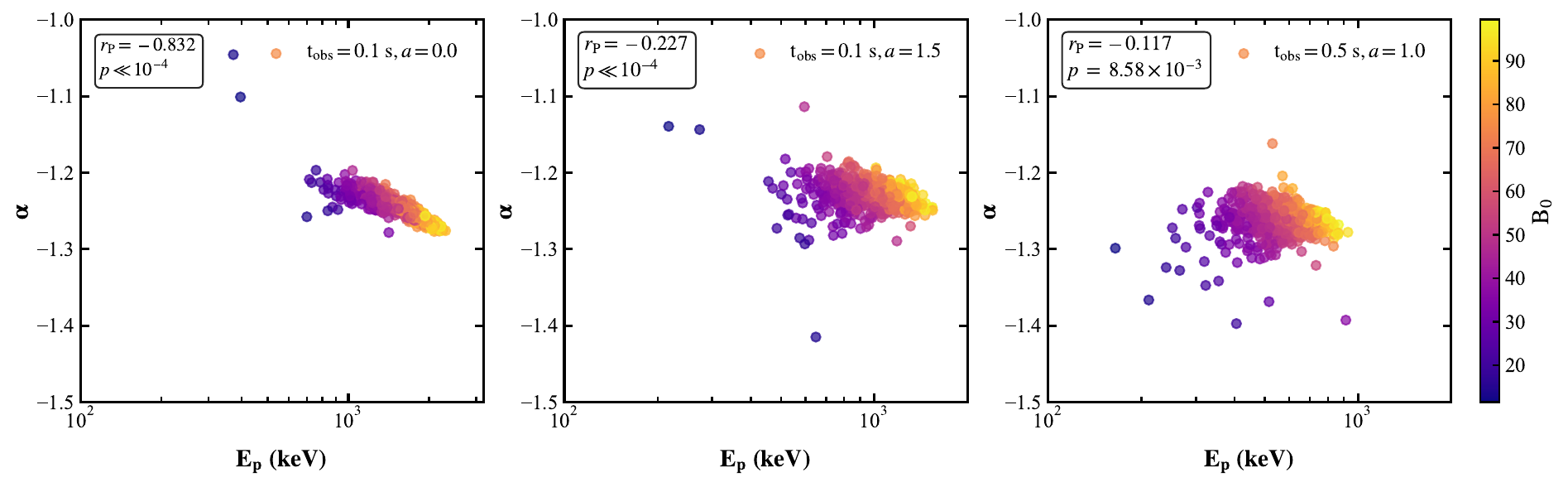}
\end{minipage}
\caption{Relation between the low-energy spectral index $\alpha$ and the peak energy $E_p$, corresponding to the models in Section \ref{section4.4}.}
\label{fig12}
\end{figure}

Figure \ref{fig12} shows the corresponding $\alpha$--$E_p$ relation for the SSC included cases. The scatter points in all three panels are concentrated within a relatively narrow range of $\alpha \approx -1.2$ to $-1.3$, indicating that SSC cooling hardens the low-energy spectral index compared with the pure synchrotron fast-cooling cases. The $\alpha$--$E_p$ correlation, however, is not uniform across the three panels: it remains relatively clear for $t_{\rm obs}=0.1$~s and $a=0.0$ ($r_{\rm P}=-0.832$), but becomes weak for $t_{\rm obs}=0.1$~s and $a=1.5$ ($r_{\rm P}=-0.227$) and for $t_{\rm obs}=0.5$~s and $a=1.0$ ($r_{\rm P}=-0.117$). Thus, although larger $B_0$ generally corresponds to higher $E_p$, the associated variation in $\alpha$ is modest. Overall, SSC cooling can mitigate the ``too soft'' problem to some extent, but the hardening is still insufficient to systematically shift $\alpha$ toward the observed peak near $-1$.

\subsection{Random Sampling Simulation}
\label{section4.5}

In this section, we perform Monte Carlo simulations to model the source evolution under two regimes: (1) synchrotron and adiabatic cooling, and (2) a combined synchrotron, adiabatic, and SSC cooling scenario. We fix the injection rate at $Q_0 = 10^{58}$ and the injection spectral index at $p = 2.8$, while the other key physical parameters are generated through random sampling. For the synchrotron + adiabatic cooling case, the observer time $t_{\rm obs}$ is sampled over the range $0.1$--$3$~s. For the simulations including SSC cooling, in order to keep the computational cost tractable, we restrict the sampling range to $t_{\rm obs}=0.01$--$0.5$ s. The initial magnetic field strength $B_0$, the emission start radius $R_0$, and the bulk Lorentz factor $\Gamma$ are sampled from Gaussian distributions in logarithmic space$\log_{10}(B_0) \sim \mathcal{N}(\mu = 1.8, \sigma = 0.2)$, $\log_{10}(R_0) \sim \mathcal{N}(\mu = 15.0, \sigma = 0.3)$, and $\log_{10}(\Gamma) \sim \mathcal{N}(\mu = 2.5, \sigma = 0.1)$. The magnetic field decay index $a$ is sampled from a normal distribution $a \sim \mathcal{N}(\mu = 1.0, \sigma = 0.5)$ and truncated within the physically permissible range $a \in [0, 2]$. To directly align the simulated spectral peaks with observations, we randomly sample $E_p$ from the Fermi-GBM catalog and inversely solve for the minimum injection Lorentz factor $\gamma'_m$ to ensure the synchrotron characteristic peak satisfies the given $E_p$, using the relation $\gamma'_m = [4\pi m_e c (1+z) E_p / (3 h q_e B' \Gamma)]^{1/2}$.

\begin{figure}[htbp]
\centering
\begin{minipage}[t]{0.5\textwidth}
\centering
\includegraphics[width=\textwidth]{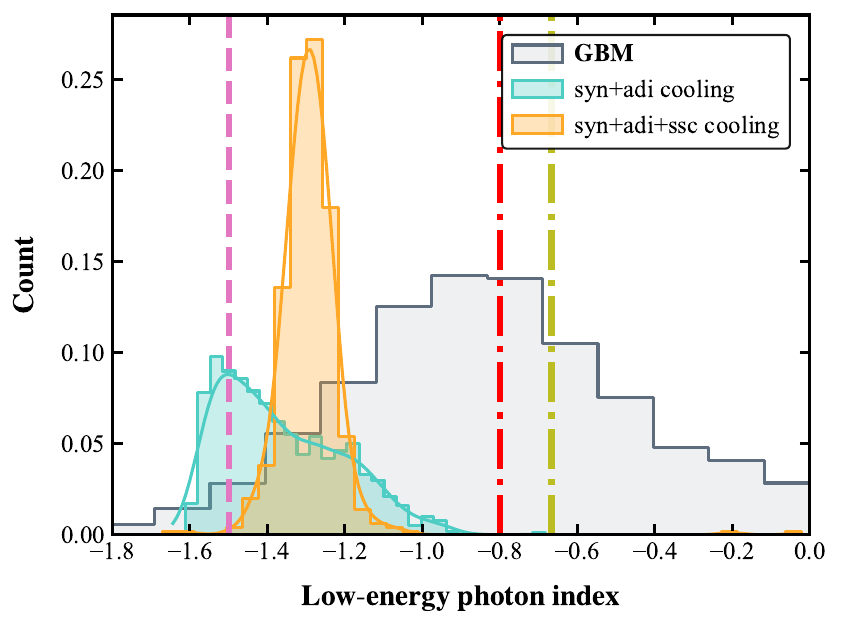}
\end{minipage}
\caption{Comparison of the Band low-energy photon-index $\alpha$ distributions between the \textit{Fermi}/GBM sample (gray) \citep{2014ApJS..211...12G}, the random-sampling simulation with synchrotron + adiabatic cooling only (cyan), and the random-sampling simulation including synchrotron + adiabatic + SSC cooling (orange) in Section~\ref{section4.5}. The magenta dashed and olive dash-dotted vertical lines mark $\alpha=-3/2$ and $\alpha=-2/3$, respectively, while the red dash-dotted line marks the peak location of the observed GBM distribution.
\label{fig13}}
\end{figure}

\begin{figure}[htbp]
\centering
\begin{minipage}[t]{0.6\textwidth}
\centering
\includegraphics[width=\textwidth]{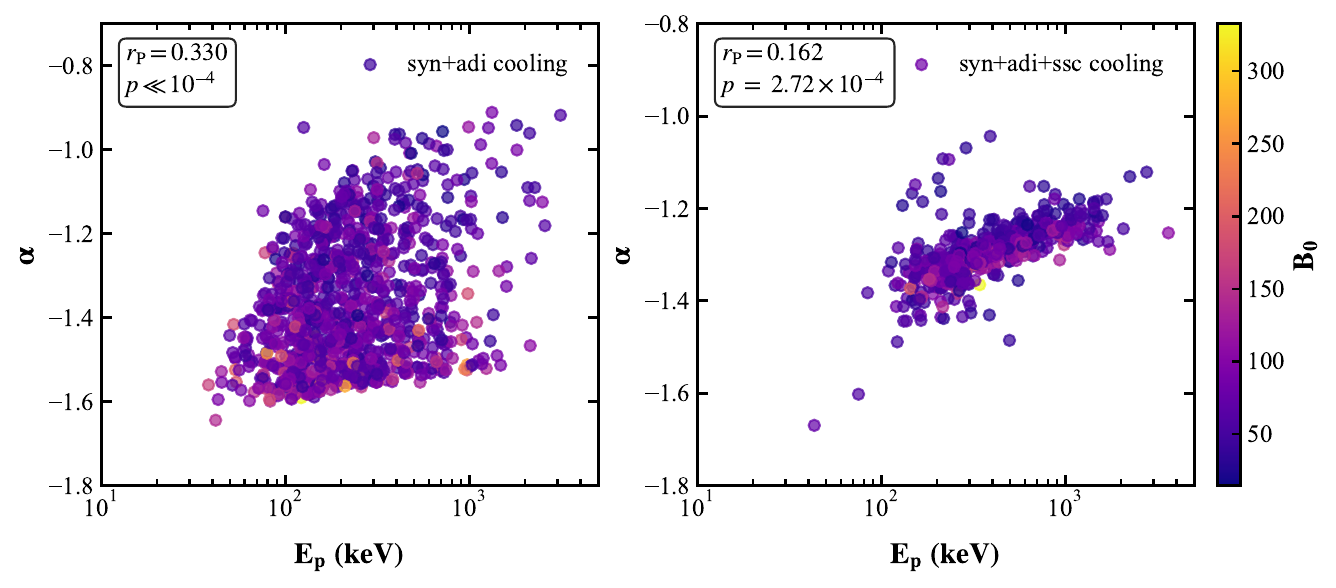}
\end{minipage}
\caption{
Relation between the low-energy spectral index $\alpha$ and the peak energy $E_p$, corresponding to the models in Section \ref{section4.5}.
\label{fig14}}
\end{figure}

Figure \ref{fig13} directly compares the low-energy spectral index distributions obtained from the Monte Carlo simulations with the \textit{Fermi}/GBM sample. The gray histogram represents the observed GBM distribution, which peaks around $\alpha \approx -0.8$ and extends toward the hard end. The cyan histogram shows the results for the synchrotron + adiabatic-cooling case. This distribution is systematically softer overall and is mainly concentrated near $\alpha \approx -1.5$. Although it exhibits a tail extending toward harder values, the fraction of events near the observed peak remains clearly insufficient, making it difficult to reproduce either the peak position or the full width of the GBM distribution. In other words, under the present sampling strategy, synchrotron radiation plus adiabatic cooling alone can produce a limited number of relatively hard events, but the population as a whole still remains close to the fast-cooling limit, $\alpha=-3/2$.

The orange histogram in Figure \ref{fig13} shows the results when SSC cooling is incorporated into the Monte Carlo simulation. Compared with the synchrotron + adiabatic cooling case, the distribution incorporating SSC cooling shifts toward harder values and becomes more concentrated around $\alpha \approx -1.3$, indicating that the introduction of SSC cooling can indeed alleviate the ``too-soft'' problem at the population level. However, even with SSC cooling included, the simulated distribution is still systematically softer than the observed GBM sample and does not fully reproduce the observed peak near $\alpha \sim -0.8$. Therefore, although the inclusion of SSC cooling improves the agreement with the data, the improvement remains partial, implying that additional physical ingredients or more complex geometric effects may still be required.

The corresponding $\alpha$--$E_p$ relations are shown in Figure \ref{fig14}. In the synchrotron + adiabatic-cooling case, the scatter points show substantial dispersion over a broad $E_p$ range and exhibit only a weak positive correlation, with a Pearson coefficient of $r_{\rm P}=0.330$. After SSC cooling is included, the scatter points shift systematically toward harder $\alpha$ values and become somewhat more concentrated, but the overall $\alpha$--$E_p$ correlation remains weak, with $r_{\rm P}=0.162$. The point color, which represents $B_0$, does not show a one-to-one correspondence with $\alpha$, indicating that under this random-sampling setup the variation of $\alpha$ is not controlled solely by $B_0$ or $E_p$. Instead, it arises from the combined effects of the cooling history, spectral curvature, and the location of the spectral peak within the finite detector band, jointly modulated by parameters such as $(R_0,\Gamma,a,t_{\rm obs})$.

\section{Summary and Discussion}
\label{section5}

We systematically investigate whether fast-cooling synchrotron emission in a decaying magnetic field can reproduce the observed statistical properties of Band-fit parameters. We solve the electron continuity equation including synchrotron, adiabatic, and SSC cooling to obtain the time-dependent electron energy distribution and the corresponding physical photon spectra; we then forward-fold the spectra through the GBM response matrices and recover $\alpha$, $\beta$, and $E_p$ with Band fits, and finally compare the resulting parameter distributions and the $\alpha$-$E_p$ correlation with the observations. Based on representative parameter studies and Monte Carlo sampling, we summarize our main findings and their physical implications as follows.

(1) Magnetic-field decay can harden the recovered $\alpha$, but the effect is not universal and is highly sensitive to observing epoch and to where the spectral peak falls within the detector band.For a representative setup with $R_0=10^{15}\,\mathrm{cm}$ and $\gamma_m=10^{5}$ (with synchrotron plus adiabatic cooling only), the $\alpha$ distribution peaks at $\simeq -1.37$ for $a=1.0$ at $t_{\rm obs}=1\,\mathrm{s}$, while a faster decay ($a=1.5$) shifts the distribution to harder values and narrows it, with a peak near $\simeq -1.23$. At later times (e.g., $t_{\rm obs}=3\,\mathrm{s}$), however, the $a=1.5$ case becomes much more dispersed and develops a soft tail with $\alpha<-1.5$ (predominantly from realizations with smaller $B_0$). This behavior indicates that once the system evolves into a weaker-field regime, the downward drift of the spectral peak and the enhanced spectral curvature project more strongly onto the finite bandpass Band fit, thereby diminishing the statistical robustness of the early-time hardening.

(2) Variations in $R_0$ and $\gamma_m$ further highlight that the catalog $\alpha$ is often an effective slope within a finite band, and should not be naively equated with the asymptotic low-energy index of the underlying spectrum.When $R_0$ is smaller (e.g., $10^{14}\,\mathrm{cm}$), the system enters the weak-field regime more rapidly for the same $t_{\rm obs}$, and the recovered $\alpha$ distribution becomes systematically softer and substantially broader (typically spanning $\alpha\sim -2.0$ to $-1.6$). When $\gamma_m$ is reduced (e.g., $\gamma_m=10^4$), the scaling $E_p \propto \Gamma B \gamma_m^2$ makes the spectral peak more likely to fall within or near the fitted band, so the fit samples the high-curvature region around the peak more frequently; as a result, the recovered $\alpha$ is driven toward softer values (approaching a concentration around $\alpha\simeq -2$). These trends show that the distribution of $\alpha$ is shaped not only by the cooling physics, but also by a coupled set of projection effects: the peak position relative to the band, the intrinsic spectral curvature, and how an empirical function absorbs that curvature. Consequently, any assessment of whether synchrotron models can account for the catalog $\alpha$ distribution must be carried out within a forward-folding framework and a catalog-consistent fitting procedure.

(3) Including SSC cooling produces non-negligible shifts in $\alpha$, but does not provide a robust route to a stable peak near $\alpha \approx -1$ across our parameter space. In representative configurations incorporating SSC cooling, the additional cooling modifies the total cooling rate and the spectral curvature, thereby altering the recovered $\alpha$ distribution. In the setups explored here, however, $\alpha$ remains concentrated in a relatively narrow range, typically $\alpha \simeq -1.2$ to $-1.3$, with limited additional hardening. In this sense, SSC cooling acts more like a partial remedy for the ``too-soft'' problem rather than a broadly applicable mechanism that robustly yields $\alpha \approx -1$ across a large volume of the parameter space.

(4) In Monte Carlo samples designed to better mimic the observed population, the synchrotron + adiabatic-cooling model remains systematically too soft relative to the GBM catalog. The observed GBM $\alpha$ distribution peaks near $\alpha \approx -0.8$ and extends to harder values ($\alpha > -1$), whereas our Monte Carlo samples in the synchrotron + adiabatic-cooling framework are dominated by $\alpha \approx -1.5$. Although a tail reaching $\alpha \gtrsim -1$ is present, the fraction of events near the observed peak and at the hard end is still clearly insufficient. After SSC cooling is included in the Monte Carlo simulations, the $\alpha$ distribution shifts systematically toward harder values and becomes more concentrated around $\alpha \approx -1.3$, indicating that SSC cooling can alleviate the ``too-soft'' problem at the population level. However, the SSC-inclusive distribution is still softer than the observed GBM sample and does not fully reproduce the observed peak near $\alpha \approx -0.8$ or the full width of the observational distribution.

In summary, a decaying magnetic field can harden $\alpha$ under certain conditions. However, synchrotron plus adiabatic and SSC cooling  remains insufficient to account for the GBM $\alpha$ distribution in a statistical sense, implying that additional physical ingredients and geometric observational effects are required \citep{2009ApJ...690L..10Z,2016ApJ...825...97U,2017ApJ...841L..15G,2024ApJ...963L..30U}.

\cite{2015MNRAS.451.1511B} demonstrated with rigorous forward-folding tests that even for synchrotron emission, the inferred $\alpha$ distribution can be strongly affected by where $E_p$ lies in the detector band and by the curvature of the underlying spectrum. Therefore, the question of whether synchrotron emission can explain the GBM $\alpha$ distribution must be addressed within a catalog-consistent fitting pipeline. Our work extends this approach to a broader class of synchrotron models with magnetic-field decay, and shows that while field decay can push fast-cooling spectra toward harder $\alpha$, it still does not fully cover the GBM peak and hard-end fraction. The contribution of this study is thus not to present a small number of hand-picked hard examples, but to provide falsifiable constraints on how far magnetic-field decay can move the catalog-recovered $\alpha$ distribution, and under what conditions the mechanism fails.

Our results also connect naturally to earlier work emphasizing changes in the dominant cooling channel. \cite{2018ApJS..234....3G} quantified, with a numerical treatment, the relative roles of synchrotron, SSC, and adiabatic cooling, and argued that low-energy hardening may arise in three regimes: SSC-dominated cooling, a declining synchrotron cooling rate with time, and adiabatic-dominated cooling in the extreme high-$\Gamma$ limit. In internal-shock scenarios where SSC dominates, they found that the low-energy spectral index can fall in the range $-2<\alpha<-1$, but also noted that pushing $\alpha$ to $\gtrsim -1.1$ can require very large energy ratios $L_e/L_B$ (up to $\gtrsim 10^{4}$) and is often associated with Klein–Nishina suppression. This is consistent with our finding that including SSC can cluster $\alpha$ around $-1.3$ to $-1.2$ but does not robustly stabilize $\alpha\approx -1$ over a broad, conservatively sampled parameter space, suggesting that reaching the hardest values may require more restrictive conditions that occupy a smaller fraction of the population. \cite{2018ApJS..234....3G} further pointed out that in magnetically dominated prompt scenarios, a rapid post-dissipation decay of the magnetic field can reduce the synchrotron cooling rate with time and yield harder low-energy spectra, an effect that our time-dependent calculations show to be most prominent at early epochs but less statistically stable at later times as the peak drifts and curvature increases \citep{2015A&A...583A.129Y,2019A&A...629A..69B}.

A complementary set of studies suggests that magnetic-field decay alone may be insufficient unless coupled to non-stationary injection histories. \cite{2021Galax...9...68W} considered internal-shock evolution with a staged magnetic-field behavior (approximately constant followed by a $t^{-1}$ decay), which can yield $-3/2<\alpha<-2/3$, and argued that allowing the electron injection rate to rise with time can further harden the spectrum toward $\alpha \sim -2/3$. \cite{2020ApJ...893L..14L}, motivated by reconnection-powered prompt emission, emphasized that the canonical fast-cooling result implicitly assumes a nearly steady injection; if $Q(t)$ varies (in particular, increases), the low-energy electron pile-up can depart from the standard $\gamma^{-2}$ form, hardening the low-energy spectral slope toward the commonly observed $\alpha\approx -1$. Combined with our Monte Carlo results, these arguments suggest a clearer ``division of labor": magnetic-field decay and  SSC can move $\alpha$ toward harder values, but they do not robustly shift the recovered distribution away from the canonical fast-cooling value ($\alpha=-3/2$) toward the observed GBM peak near $\alpha\sim -1$. A natural next step is to incorporate $Q(t)$ (e.g., $Q\propto t^{q}$) together with $B'\propto R^{-a}$, while retaining the same forward-folding and Band-recovery pipeline, to disentangle whether the hardening is driven primarily by a reduced cooling rate or by an altered injection history.

Persistent energization provides another plausible route. \cite{2018ApJ...853...43X} proposed adiabatic stochastic acceleration (ASA) as a mechanism that can continuously supply energy to electrons over a finite range and thereby counteract synchrotron cooling, reshape the low-energy electron distribution, and yield harder low-energy spectral indices near $\alpha\approx -1$. They also noted that magnetic-field decay can, in principle, harden fast-cooling spectra. In light of our catalog-level constraints, ASA or other sustained-heating scenarios may be more naturally interpreted as complementary ingredients that enlarge the parameter volume producing hard $\alpha$ and improve statistical robustness, rather than as an exclusive alternative to magnetic-field decay.

Finally, geometric effects and observational selection may play a non-negligible role.
\cite{2025ApJ...986..106G} showed, through Monte Carlo population syntheses, that distributions of viewing angles,
Doppler-factor variations, and equal-arrival-time surface integration can (in their simulations) soften the
recovered $\alpha$ and broaden its distribution even if the underlying microphysics is unchanged.
The strong dispersion in our $\alpha$-$E_p$ plane and the absence of a one-to-one mapping between $\alpha$ and any
single model parameter suggest that incorporating viewing-angle effects and multi-zone superposition within the
same forward-folding framework could further modify the width and the hard-end fraction (e.g., $\alpha>-1$) of the
recovered distributions. Therefore, given that our model remains systematically too soft and underproduces the hard end, coupling the radiative calculations to a self-consistent geometric/observational mapping is a high-priority extension, alongside more complex time-dependent injection \citep{1998MNRAS.296..275D,2014A&A...568A..45B} and sustained-heating prescriptions \citep{2009ApJ...705.1714A,2015MNRAS.454.2242A}.

Overall, within a catalog-consistent forward-folding and Band-recovery framework, we find that magnetic-field decay and SSC can harden the low-energy slope of fast-cooling synchrotron emission in some regimes, but they do not by themselves reproduce the peak and width of the GBM $\alpha$ distribution statistically. A more realistic picture likely requires magnetic-field decay to operate in concert with non-stationary injection, sustained acceleration, and geometric effects, with their relative contributions quantified through the same forward-folding and uniform-fitting pipeline.  Moreover, within the class of synchrotron models with decaying fields explored here, our simulations account for at most somewhat less than half of the catalog-recovered spectra under our adopted matching criteria. This suggests that a substantial fraction of GRBs may involve additional emission components, such as photospheric emission including sub-photospheric dissipation \citep{2010MNRAS.407.1033B,2011MNRAS.415.3693R}, as well as more complex geometrical effects, such as structured jets \citep{2015MNRAS.450.3549S,2020A&A...641A..61A}.

\begin{acknowledgments}
We acknowledge the use of the public data from the Fermi data archives. This work is supported by the National Natural Science Foundation of China (NSFC 12233006). Dr. Shan Chang acknowledges support from the National Natural Science Foundation of China 12103046 and the Xingdian Talent Support Plan - Youth Project. This work is also supported by the Postdoctoral Fellowship Program of CPSF under Grant Number GZC20252090 and the Yunnan University graduate research innovation fund project KC-24248607.
\end{acknowledgments}

\bibliography{references}{}

\bibliographystyle{aasjournalv7}
\end{document}